\documentclass[aip,rsi,reprint,graphicx]{revtex4-1} 
\usepackage{amsmath,amsthm,amssymb}
\usepackage{graphicx}
\usepackage{ulem}
\usepackage{color}

\draft % marks overfull lines with a black rule on the right

\begin{document}

%\preprint{}

\title{High-$\beta$ equilibrium and ballooning stability of the low aspect 
       ratio CNT stellarator} 



\author{K.~C.~Hammond}
\email[]{kch2124@columbia.edu}
%\thanks{}
\affiliation{Columbia University, New York, NY 10027}

\author{S.~A.~Lazerson}
\affiliation{Princeton Plasma Physics Laboratory, Princeton, NJ 08536}

\author{F.~A.~Volpe}
\email[]{fvolpe@columbia.edu}
\affiliation{Columbia University, New York, NY 10027}


\date{\today}

\begin{abstract}
The existence and ballooning-stability of low aspect ratio stellarator
equilibria is predicted for the Columbia Neutral Torus (CNT) with the aid of 3D 
numerical tools.
In addition to having a low aspect ratio, CNT is characterized by a low 
magnetic field and
small plasma volume. Also, highly overdense plasmas were recently 
heated in CNT by means of microwaves. These characteristics 
suggest that CNT might attain relatively high values
of plasma beta and thus be of use in the experimental study of stellarator 
stability to high-beta instabilities such as ballooning modes. As a first step
in that direction, here the ballooning stability limit is found numerically.
Depending on the particular magnetic configuration we expect
volume-averaged $\beta$ limits in the range 0.9-3.0\%, and possibly higher,
and observe indications of a second region of ballooning stability.
As the aspect ratio is reduced, stability is found to increase in some 
configurations and decrease in others.
Energy-balance estimates using stellarator scaling laws indicate that 
the lower $\beta$ limit may be attainable with overdense 
heating at powers of of 40 to 100 kW.
The present study serves the additional purpose of testing VMEC and other 
stellarator codes at high values of $\beta$ and at low
aspect ratios. For this reason, the study was carried out both for free
boundary, for maximum fidelity to experiment, as well as with a fixed boundary,
as a numerical test.
\end{abstract}


\maketitle %\maketitle must follow title, authors, abstract and \pacs

% References should be done using the \cite and \label commands
\section{Introduction}
\label{sect:intro}

To date, the highest plasma beta among stellarators---about 5\%---have been 
obtained in W7-AS
\cite{weller2003} and in LHD \cite{sakakibara2015}. Neither plasma was found to 
be unstable, implying even higher stability limits in those devices. Therefore, 
to date analytical and numerical investigations of ballooning 
modes \cite{cuthbert1998,hegna1998,hudson2003,hudson2004,rafiq2010} and other 
high-$\beta$ instabilities \cite{nielson2000}
have only received partial validation by experiment:
experiments confirmed certain values of $\beta$ to be stable, but could not 
verify whether even higher values were unstable\cite{weller2001,weller2003}. 
Experimental access to the
stellarator $\beta$ limit and experimental characterization of instabilities
would finally enable comparison with theory and improve our understanding and 
predictive capability. Access to higher $\beta$ (and, yet, 
stability) could also lead to more compact and efficient 
stellarator reactor designs, currently assuming volume-averaged beta 
$\langle\beta\rangle=$ 3-6\% \cite{beidler2001,sagara2010}. In fact, the main 
optimization criterion in the HELIAS reactor is to maintain a stability 
limit $\langle\beta\rangle>4\%$ while reducing the Pfirsch-Schl\"uter
currents and Shafranov shift \cite{beidler2001}.

\begin{figure}
    \includegraphics[width=0.5\textwidth]{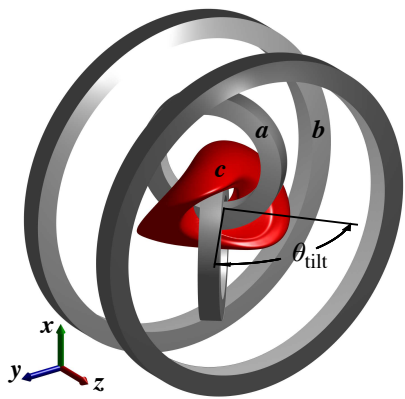}
    \caption{Schematic of the CNT coil configuration. 
            (a) interlocked (IL) coils;
            (b) poloidal field (PF) coils;
            (c) last closed flux surface.
            Adapted with permission from Ref.~\cite{hammond2016}.}
    \label{fig:schematic}
\end{figure}

In the present work, it is argued that the Columbia Neutral Torus (CNT) 
stellarator at Columbia 
University (Fig.~\ref{fig:schematic}) is uniquely well-suited for this research.
The device, originally constructed to study non-neutral and pure-electron
plasmas \cite{pedersen2004,pedersen_pop2006,kremer2006,sarasola2012}, 
has since been repurposed
to investigate quasi-neutral plasmas, and has addressed issues relevant to
magnetic fusion energy such as error-field diagnosis \cite{hammond2016} and
processing of stellarator images \cite{hammond_rsi2016}.
From the point of view of high $\beta$ stability, CNT
is attractive for two main reasons: (1) it could reach
high values of $\beta$ by deploying relatively small amounts of heating power
and (2) its stability limit is 
expected to be lower and thus more easily accessible than in other devices.

Regarding the first point, the CNT magnetic field is low: in general $B < 0.3$
T, but $B < 0.1$ T was adopted for this work. As a result, the magnetic pressure
$B^2/2\mu_0$ is very low, and more amenable plasma pressures (2500 times lower
than in a 5 T reactor, if not smaller) suffice to reach high $\beta$. The need
for high plasma pressure will require heating at high density. In this
regard, overdense plasma heating, at densities in excess of four times the 
cutoff density, was recently observed in CNT by injecting 10 kW microwaves at
2.45 GHz. Increasing the heating power would result in high power densities
in the small CNT plasma ($V \approx 0.1 \text{m}^3$). On the other hand, the 
small size of CNT is co-responsible for poor energy confinement. Even so, 
scaling-law calculations to be presented in this paper suggest that
hundreds of kW of microwave power might be 
sufficient for high $\beta$.

Regarding the second point, CNT is a classical, non-optimized stellarator.
In particular, it was not optimized for high stability, making its stability 
limit lower and easier to access. An interesting competing effect
might arise from the CNT low aspect ratio, $A \ge 1.9$. This characteristic,
relatively under-explored in stellarators, increased the stability limit in 
spherical tokamaks and favored the achievement of higher $\beta$ compared
to tokamaks \cite{}. It is interesting to verify whether the low 
aspect ratio has a similar beneficial effect on stellarator stability,
although evidence presented below suggests this not to be the case, at least
not for CNT.

This paper describes a numerical investigation of high-$\beta$ 
equilibria that are attainable in the CNT configuration. Equilibria are 
calculated using the VMEC code,\cite{hirshman1983} which 
solves the ideal MHD equations for input vacuum fields and profile functions. 
They are then evaluated for stability using COBRAVMEC,\cite{sanchez2000}
which determines ballooning growth rates at various locations 
in the plasma. The structure is as follows: 
Sec.~\ref{sect:fixedBdry} briefly overviews considerations for calculating 
VMEC equilibria in the CNT configuration and presents some 
fixed-boundary results. Sec.~\ref{sect:inputParams} reviews the assumptions
used for calculations of equilibrium parameters, bootstrap current, and 
stability. The methods for the calculations are then described in 
Sec.~\ref{sect:equilStabCalc}. Sec.~\ref{sect:highBeta_results} presents the
main free-boundary equilibrium and stability results. Sec.~\ref{sect:scaling} 
describes scaling-law calculations to predict how much heating
power will be necessary to attain the equilibria in 
Sec.~\ref{sect:highBeta_results}.

\section{Fixed-boundary VMEC solutions for CNT geometry}
\label{sect:fixedBdry}

VMEC may assume a fixed or free plasma boundary depending on the purpose of the 
calculation. In fixed-boundary mode, the plasma boundary is assumed to be known
\textit{ab initio} and the full magnetic field is determined in the 
calculation without 
knowledge of the external coils. In free-boundary mode, the coil configuration
and the field that it generates are known and the plasma boundary is determined
using an energy principle.\cite{hirshman1986} Free-boundary mode was used in
this work because the shape of the plasma boundary was expected to vary with 
$\beta$ and plasma current.

\begin{figure}
    \includegraphics[width=0.5\textwidth]{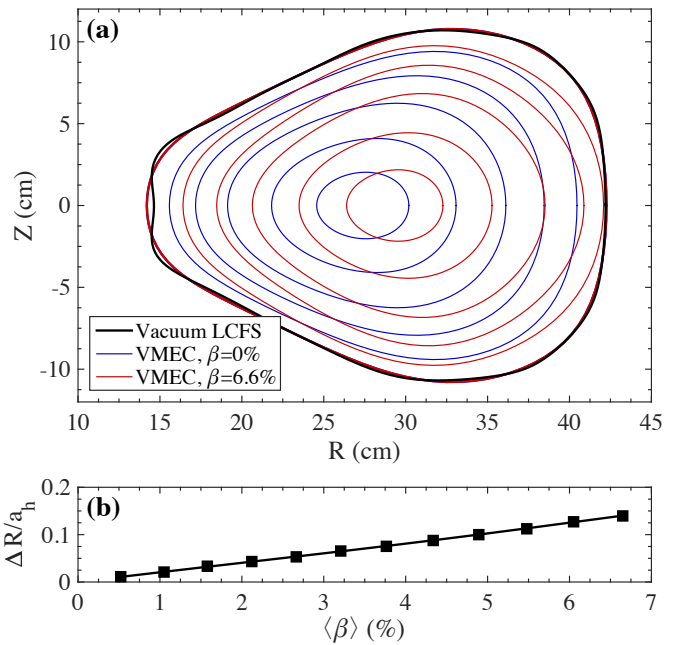}
    \caption{Results of a series of fixed-boundary simulations of CNT with 
             $\theta_\text{tilt}=78^\circ$, $I_{IL}/I_{PF}=2.5$, 
             $I_\text{p}=0$, and a centrally peaked pressure profile as shown
             in Fig.~\ref{fig:presProf}. 
             (a) Flux surface comparison between calculations with 
                 $\langle\beta\rangle=0\%$ and $\langle\beta\rangle=6.6\%$,
                 with the vacuum last closed flux surface (LCFS) as a reference;
             (b) Relative Shafranov shift as defined in the text.}
    \label{fig:fixedBdry}
\end{figure}

As an initial test of concept, however, a number of high-$\beta$ calculations
were performed in fixed-boundary mode for CNT-like configurations. 
%(CNT's magnetic
%configurations may be paramatrized by (1) the tilt angle, $\theta_\text{tilt}$
%between the interlocked (IL) coils, which may be set to $64^\circ$, $78^\circ$,
%or $88^\circ$, and (2) the ratio of current, $I_{IL}/I_{PF}$ between the IL
%and poloidal field (PF) coils.) 
One example is shown in Fig.~\ref{fig:fixedBdry}. In this case, the boundary 
was set to conform to the last closed flux surface (LCFS) obtained from vacuum 
field line calculations with $\theta_\text{tilt}=78^\circ$ and 
$I_{IL}/I_{PF} = 2.5$. Here $\theta_\text{tilt}$ denotes the tilt angle between
the interlocked (IL) coils (Fig.~\ref{fig:schematic}). This angle can be set to
$64^\circ$, $78^\circ$, or $88^\circ$. $I_{IL}/I_{PF}$ is the ratio of currents
flowing in the IL and poloidal field (PF) coils.
For this choice of $\theta_\text{tilt}$ and $I_{IL}/I_{PF}$, fixed-boundary 
equilibria were obtainable for $\langle\beta\rangle$ up to
$6.6\%$ (Fig.~\ref{fig:fixedBdry}). The relative Shafranov shift 
$\Delta R/a_h$, defined according to
Ref.~\cite{weller2006} as the horizontal shift $\Delta R$ in the 
magnetic axis over the horizontal minor radius $a_h$, is seen to
increase linearly with $\beta$, as expected (Fig.~\ref{fig:fixedBdry}b).
While these results are not particularly relevant to future comparisons with
experiment, they are useful nonetheless as a verification that VMEC yields
reasonable results at high beta and low aspect ratio, comparable to 
calculations made for a spherical stellarator concept in 
Ref.~\cite{moroz1996}.

\section{Input parameters}
\label{sect:inputParams}

CNT's field strength for electron cyclotron heated (ECH) plasmas, and possibly
electron Bernstein wave heated (EBWH) plasmas, is constrained by 
the requirement for $|\mathbf{B}| = 0.0875$ T for heating at the first electron
cyclotron harmonic or $|\mathbf{B}| = 0.0437$ T for the second harmonic at 
2.45 GHz. With $|\mathbf{B}|$ fixed, the only two degrees of freedom 
controlling the vacuum-field configuration are $\theta_\text{tilt}$ and 
$I_{IL}/I_{PF}$.

The size of the free boundary is determined by the 
total enclosed magnetic flux. This was chosen to approximately equal the 
flux enclosed in the vacuum-field LCFS for the respective configuration.

\begin{figure}
    \includegraphics[width=0.4\textwidth]{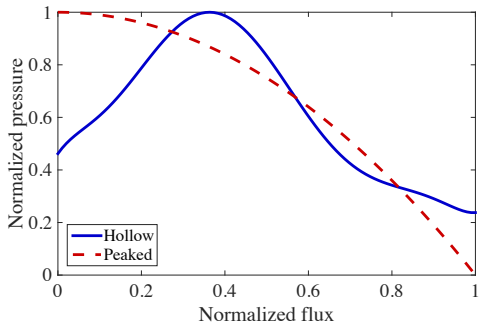}
    \caption{Normalized profiles of pressure used as inputs for the 
             calculations in this paper. During the $\beta$ scans, each of
             these profiles was scaled according to
             the desired $\beta$ value.}
    \label{fig:presProf}
\end{figure}

Pressure profiles were assumed to have one of the two functional forms shown in 
Fig.~\ref{fig:presProf}. The hollow profile corresponds to a typical 
Langmuir probe measurements of ECH plasmas in CNT. The peaked profile shown
was also considered for comparison, because peaked profiles cannot be 
ruled out from future experiments.

As CNT does not have a solenoid transformer and does not deploy any form of 
current drive (EC, EBW, or other), the toroidal current density was 
assumed to be equal to the bootstrap current density, $J_\text{bs}$. Profiles of
$J_\text{bs}$ were calculated for each equilibrium using the BOOTSJ 
code, which evaluates bootstrap currents in nonaxisymmetric
plasma configurations using the drift kinetic equation.\cite{shaing1989} 

Since BOOTSJ requires the electron density and the electron and ion 
temperatures as inputs, these quantities were estimated as follows. For 
equilibria with the \textit{hollow} (experimentally obtained) pressure 
profiles, the electron temperature profile was chosen to resemble the 
corresponding temperature measurements. For equilibria with the \textit{peaked} 
pressure profiles, since no corresponding 
experimental data are available, temperature profiles were assumed to be flat.
As CNT does not employ any direct ion heating mechanisms, the
ion temperature was assumed to be 0.3 times the electron temperature at all
locations. 
The electron temperature was constrained to not exceed 30 eV 
(hence, increases in $\beta$ were mostly driven by increases in density). The 
desire for low temperature and high density arises from the energy confinement 
scaling (Eq.~\ref{eqn:tauE}). We impose a lower limit of 30 eV to avoid 
excessive radiative losses associated with peak radiation from hydrogen 
isotopes, other working gases such as noble gases,
and common impurities such as oxygen and carbon.\cite{post1977}

%Furthermore, because the
%low-$\beta$ plasmas currently produced in CNT are already overdense to 
%2.45 GHz electron cyclotron waves by factors of up to 4 (for electron 
%temperatures of $\approx$ 5 eV), further increases in $\beta$ 
%were assumed to be driven by proportional increases in temperature 
%(as opposed to increases in density). With effective charge assumed to be 1,
%the particle densities then followed from the temperatures and the plasma
%pressure profile.

\section{Methods for equilibrium and stability calculations}
\label{sect:equilStabCalc}

The maximum achievable $\beta$ for each configuration (defined by
$\theta_\text{tilt}$, $I_{IL}/I_{PF}$, field strength, and pressure profile) 
was determined using the following procedure: 

\begin{enumerate}
    \item Conduct a free-boundary VMEC calculation with $\beta = 0$ and zero
          toroidal current.
    \item Using the results of the previous step as an initial guess, determine
          a free-boundary equilibrium with the pressure incrementally increased
          in magnitude (while maintaining the profile shape in 
          Fig.~\ref{fig:presProf}).
    \item Calculate the bootstrap current profile, $dI_\text{bs}/d\psi_n$, from 
          the result of the previous step. $\psi_n$ is the normalized toroidal
          magnetic flux, which is used as a surface coordinate.
    \item Re-calculate the equilibrium in step 2, incorporating the output
          from step 3 to obtain a self-consistent result.
    \item Repeat steps 2-4 until either: (1) VMEC fails to descend robustly to
          an equilibrium solution or (2) the solution is not stable to 
          ballooning instabilities, as determined below.
\end{enumerate}

Each self-consistent equilibrium was evaluated for ballooning stability
by the COBRAVMEC code, which computed growth rates on a grid of 1620 locations 
(15 toroidal $\times$ 18 poloidal $\times$ 6 radial) throughout 
the plasma volume. The maximum among these values was defined as  
$\gamma_\text{max}$. $\gamma_\text{max}$ was recorded for each equilibrium
(\textit{i.e.}, for each $\beta$ increment) such that $\gamma_\text{max}$ 
could be plotted as a function of $\beta$. The maximum ballooning-stable $\beta$
was then defined as the highest value of $\langle\beta\rangle$ for which 
$\gamma_\text{max} < 0$. This was obtained through 
interpolation of
$\gamma_\text{max}(\beta)$ (as in, for example, Fig.~\ref{fig:qty_vs_beta}c).

This procedure was carried out in three different plasma parameter regimes. 
The first used the experimental hollow pressure profile 
(Fig.~\ref{fig:presProf}) and a magnetic field strength $|\mathbf{B}|$ 
appropriate for first-harmonic ECH (denoted hereafter by $B\approx0.08$ T).
The second used the same pressure profile but half the field strength as 
is appropriate for second-harmonic ECH (denoted by $B\approx0.04$ T).
The third used the peaked pressure profile and $B\approx0.08$ T.

%In executing the above calculations, it was observed that VMEC needed to be
%operated at a low spatial resolution in order to descend robustly to 
%equilibrium solutions for the CNT configuration. VMEC models the plasma as 
%a set of discrete, nested flux surfaces whose geometry is parameterized
%by Fourier series in the poloidal angle $\theta$ and toroidal angle $\zeta$
%as follows:
%
%\begin{align}
%    R_s(\theta, \zeta) = 
%        \sum_{m=0}^M \sum_{n=-N}^N R_{mn} \cos(m\theta - pn\zeta) \\
%    Z_s(\theta, \zeta) = 
%        \sum_{m=0}^M \sum_{n=-N}^N Z_{mn} \sin(m\theta - pn\zeta) 
%\end{align}
%
%\noindent Here, $R$ and $Z$ are the cylindrical coordinates of a point on a 
%flux surface $s$ parametrized by $\theta$ and $\zeta$. Stellarator symmetry is 
%assumed with $p$ field periods. For the simulations described in this paper,
%$M$ was restricted to 5 and the plasma was discretized into no more than 30 
%surfaces, the latter of which is lower than typical for VMEC. These
%restrictions inevitably obscure some of the fine geometric features that 
%characterize CNT's vacuum flux surfaces. 
%$N$ was set 
%to 5 -- not due to limitations for robust descent, which were found to not be
%as stringent for $N$, but because this was sufficient for characterizing 
%the toroidal variations in surface geometry. 

\section{Free-boundary and stability results}
\label{sect:highBeta_results}

\begin{figure}
    \includegraphics[width=0.5\textwidth]{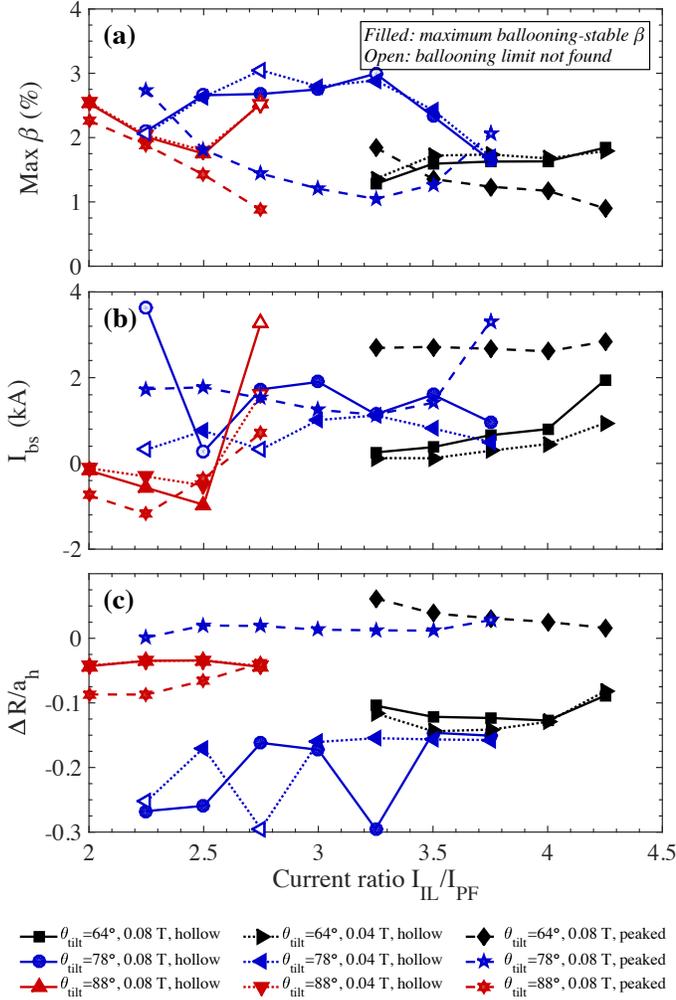}
    \caption{(a) Highest volume-averaged $\beta$,
             (b) bootstrap current, and
             (c) relative Shafranov shift
             as functions of the coil current ratio ($I_{IL}/I_{PF}$) and
             $\theta_\text{tilt}$ for the three different configurations of
             pressure and field strength discussed in the text. 
             Here, ``highest $\beta$'' refers to the highest 
             value before the plasma becomes ballooning-unstable (filled
             symbols) or the highest value for which it was possible to 
             compute a stable equilibrium, without however encountering the
             ballooning stability limit yet (open symbols; see also 
             Fig.~\ref{fig:qty_vs_beta}c). 
             Some markers overlap one another due to similar results.} 
     \label{fig:maxBeta}
\end{figure}

Maximum volume-averaged $\beta$ values for attainable tilt angles and current
ratios are plotted in Fig.~\ref{fig:maxBeta} alongside other quantities of 
interest. The highest volume-averaged $\beta$ not vulnerable to ballooning
instability was 3.0\%. This was obtained in 
two configurations, one with $B\approx 0.04$ T and $I_{IL}/I_{PF} = 2.75$; 
the other with $B\approx 0.08$ T and $I_{IL}/I_{PF} = 3.25$. Both had
$\theta_\text{tilt}=78^\circ$ and the hollow pressure profile.
The latter is expected to be easier to attain experimentally due to the 
favorable scaling of confinement time with $B$ (Eq.~\ref{eqn:tauE}) and will
be referred to hereafter as the high-$\beta$ configuration.
The highest stable $\beta$ values attained
in the other two tilt angles were 2.5\% for 
$\theta_\text{tilt} = 88^\circ$ and 1.8\% for $\theta_\text{tilt} = 64^\circ$.
Open markers in
Fig.~\ref{fig:maxBeta} represent configurations in which the VMEC code did not
find equilibria that were ballooning-unstable. In other words,
the actual $\beta$ limit for the configurations denoted by open symbols could
be even higher than shown in Fig.~\ref{fig:maxBeta}. One such configuration is
the high-$\beta$ configuration described above: its $\beta$ limit could in 
fact be higher than 3\%.

\begin{figure}
    \includegraphics[width=0.5\textwidth]{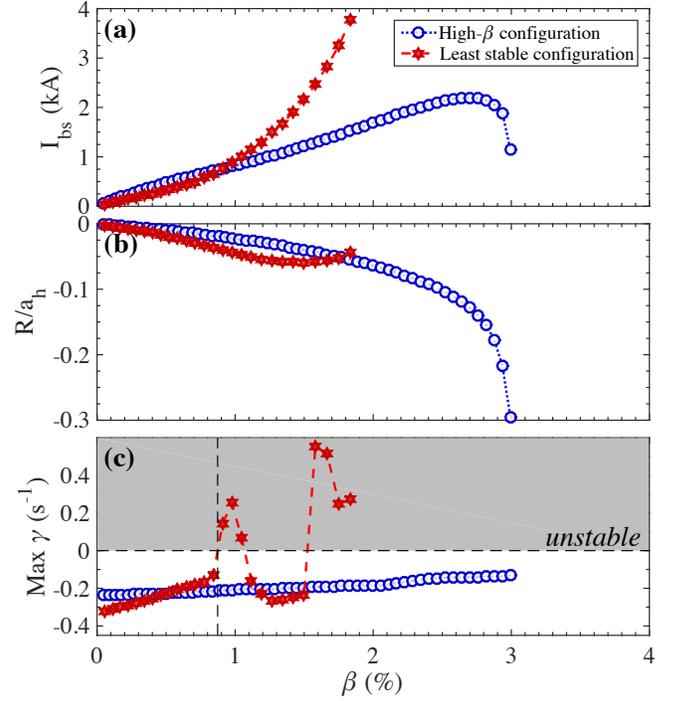}
    \caption{Evolution of key paramaters during the $\beta$ scan carried out
             for the high-$\beta$ magnetic configuration 
             ($\theta_\text{tilt} = 78^\circ$, $I_{IL}/I_{PF} = 2.75$, 
             $B = 0.04$ T, hollow pressure profile) and for the configuration
             that became ballooning-unstable at the lowest $\beta$
             ($\theta_\text{tilt} = 88^\circ$, $I_{IL}/I_{PF} = 2.75$, 
             $B = 0.08$ T, peaked pressure profile). The vertical dashed line
             indicates the value of $\beta$ at which the least stable
             configuration first becomes ballooning unstable.
             (a) bootstrap current;
             (b) relative Shafranov shift;
             (c) maximum calculated ballooning growth rate within the plasma
                 volume.
             Note that the growth rate never becomes positive 
             for the high-$\beta$ configuration.}
    \label{fig:qty_vs_beta}
\end{figure}

%The equilibria with the maximum and minimum
%ballooning-unstable $\beta$ were both obtained for
%$\theta_\text{tilt}=78^\circ$ and $\theta_\text{tilt}=88^\circ$,
%respectively. The former had the hollow profile, and
%the latter had the peaked profile.
%Evolution of some key parameters for these two configurations 
The lowest $\beta$ threshold for ballooning stability (0.9\%) was found in a 
configuration with $\theta_\text{tilt}=88^\circ$ and the peaked pressure 
profile. This will be referred to hereafter as the least stable configuration. 
The evolution of some key parameters for this and the high-$\beta$ configuration
during the $\beta$ scan are shown in Fig.~\ref{fig:qty_vs_beta}.
The nonlinearity in relative Shafranov shift $\Delta R/a_h$ 
(Fig.~\ref{fig:qty_vs_beta}b) is due to the change in the shape of the plasma
with $\beta$, resulting in a non-constant $a_h$.
%The configuration with the lowest ballooning-unstable 
%$\beta$ will be referred to hereafter as the least stable configuration.

\begin{figure}
    \includegraphics[width=0.5\textwidth]{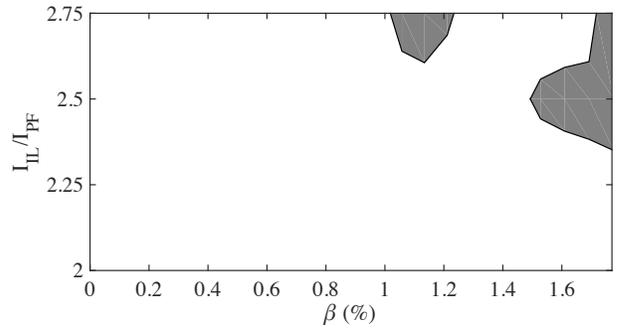}
    \caption{Contour plot of ballooning stability in 
             $(\beta, I_{IL}/I_{PF})$ parameter space for the peaked-profile
             configurations with $\theta_\text{tilt} = 88^\circ$, $B = 0.08$ T.
             The shaded regions are unstable ($\gamma_\text{max} > 0$).}
    \label{fig:stab_contours}
\end{figure}

Note that while the least stable configuration initially 
becomes unstable at $\beta=0.9\%$, it exhibits a second region of stability 
for $1.1\% < \beta < 1.5\%$ (Fig.~\ref{fig:qty_vs_beta}c), probably due to the
high bootstrap current (Fig.~\ref{fig:qty_vs_beta}a) and consequently high
shear. Second regions of 
ballooning stability have been the subject of theoretical research for both 
tokamak and stellarator configurations (for example, 
Refs.~\cite{greene1981,hudson2003,hudson2004}). Fig.~\ref{fig:stab_contours}
shows that this region could be accessed by a proper ``trajectory'' in a 
two-dimensional space spanned by heating power (roughly proportional to $\beta$
and to the pressure gradient) and coil-current ratio (controlling the 
\sout{$\iota$} profile, hence magnetic shear).

The total bootstrap-currents $I_\text{bs}$ (Fig.~\ref{fig:maxBeta}b) are low in 
comparison with the coil-currents (40 to 90 kA-turns in the IL coils). For the
configurations with hollow pressure profiles, this is partly a result of
$J_\text{bs} < 0$ near the axis and
$J_\text{bs} > 0$ near the edge. These current profiles also lead to negative 
Shafranov shifts for many of the configurations (Fig.~\ref{fig:maxBeta}c).

\begin{figure}
    \includegraphics[width=0.5\textwidth]{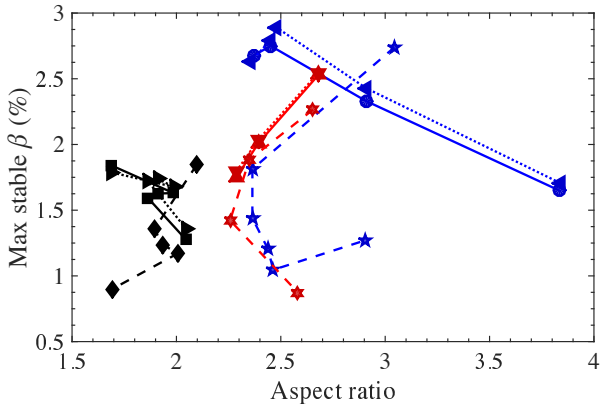}
    \caption{Maximum stable $\beta$ plotted against aspect ratio for each
             of the test configurations. Symbols are as defined in the legend 
             in Fig.~\ref{fig:maxBeta}.}
    \label{fig:beta_vs_aspect}
\end{figure}

Fig.~\ref{fig:beta_vs_aspect} shows the maximum ballooning-stable $\beta$ 
values for each configuration considered in Fig.~\ref{fig:maxBeta}, plotted 
against the aspect ratio. The figure does not show configurations for which a 
stability threshold was not found.
The maximum stable $\beta$ did not exhibit a clear 
trend---growing for some configurations and 
decreasing for others.
This is in contrast with tokamaks, where lower aspect ratios correlate with
greater stability to ballooning modes and ideal kinks \cite{gryaznevich1998}.

\begin{figure*}
    \includegraphics[width=\textwidth]{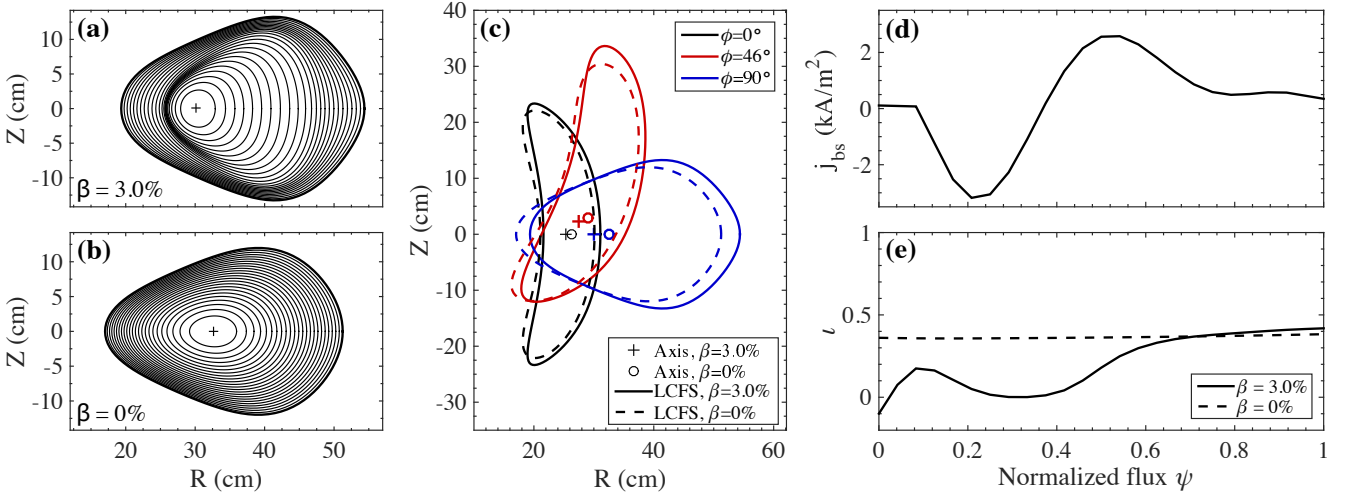}
    \caption{Properties of the high-$\beta$ equilibrium attained with 
             $I_{IL}/I_{PF}$ = 3.25, $\theta_\text{tilt} = 78^\circ$, and
             the hollow pressure profile.
             (a) Flux surfaces for the equilibrium with 
                 $\beta$ = 3.0\%.
             (b) Flux surfaces for the same configuration with $\beta$ = 0\%.
             (c) Boundaries for the equilibrium with 
                 $\beta$ = 3.0\% (solid 
                 lines) compared with boundaries calculated with $\beta = 0$
                 (dashed lines) for three different poloidal cross-sections.
             (d) Profile of toroidal current (\textit{i.e.}, the calculated 
                 bootstrap current).
             (e) Profile of rotational transform with 
                 $\beta$ = 3.0\% (solid
                 line) and $\beta$ = 0\% (dashed line).}
    \label{fig:highPerf_comp}
\end{figure*}

Characteristics of the high-$\beta$ equilibrium are shown in 
Fig.~\ref{fig:highPerf_comp}. Figs.~\ref{fig:highPerf_comp}a-c compare the
plasma geometry for $\beta=3.0\%$ to the vacuum configuration 
($\beta = 0$). Due to the sign change of the bootstrap current profile
(Fig.~\ref{fig:highPerf_comp}d),
the radial shifts of the axis and the boundary are in opposite directions
(Fig.~\ref{fig:highPerf_comp}c). The toroidal current significantly affects the 
rotational transform profile near the axis (Fig.~\ref{fig:highPerf_comp}e) but 
less so near the edge.
%Interestingly, the radial shift of the boundary (3 cm; 
%Fig.~\ref{fig:maxBeta}a) at
%$\phi=90^\circ$ is markedly greater than the shift of the axis (1 mm; 
%Fig.~\ref{fig:highPerf_comp}c). This is an effect of the sign change of the 
%current profile at an intermediate minor radius (Fig.~\ref{fig:highPerf_comp}b).

\begin{figure*}
    \includegraphics[width=\textwidth]{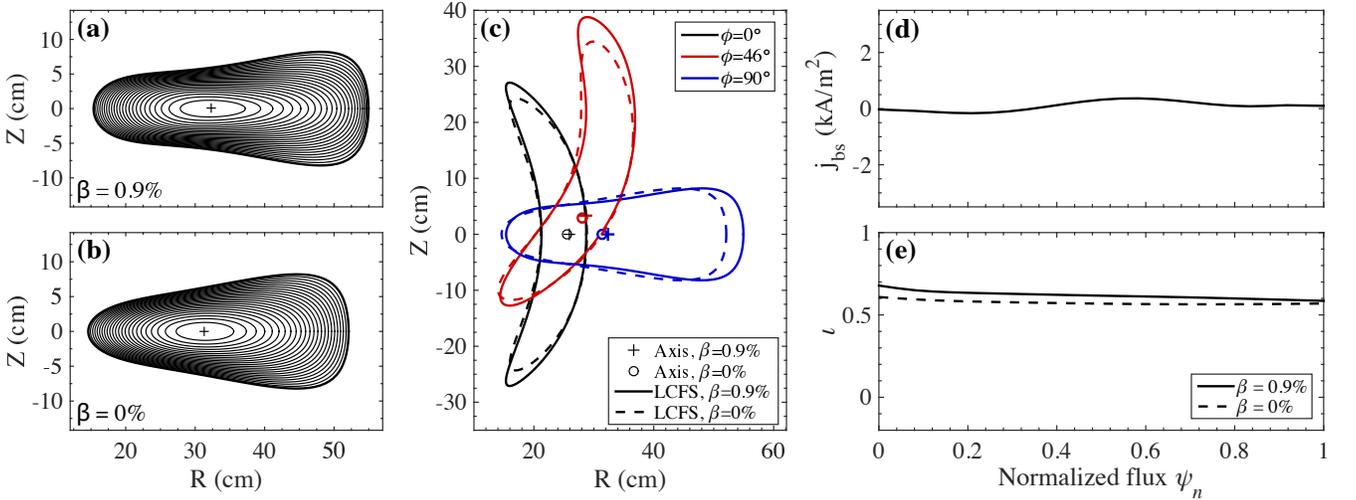}
    \caption{Like Fig.~\ref{fig:highPerf_comp}, but for the least stable 
             equilibrium ($\beta = 0.9\%$), attained with 
             $I_{IL}/I_{PF}$ = 2.75, $\theta_\text{tilt} = 88^\circ$,
             and peaked pressure profile.}
    \label{fig:unstable_comp}
\end{figure*}

Corresponding characteristics of the least stable configuration are shown in 
Fig.~\ref{fig:unstable_comp}. The differences in flux surface geometry 
(Fig.~\ref{fig:unstable_comp}a-c) and rotational transform
(Fig.~\ref{fig:unstable_comp}e) from the high-$\beta$ configuration result 
primarily
from the different $\theta_\text{tilt}$ and $I_{IL}/I_{PF}$. The lower
magnitude of the bootstrap current (Fig.~\ref{fig:unstable_comp}d), as well
as the smaller excursion of \sout{$\iota$} from its vacuum values 
(Fig.~\ref{fig:unstable_comp}e), are consistent with the lower pressure.

It should be noted that perfect stellarator symmetry was assumed for the 
calculations here for computational simplicity. However, it is known that 
CNT's coils have significant misalignments and exhibit field errors that 
break the two-field-period stellarator symmetry.\cite{hammond2016} The 
principal effect of
these errors was shown to be an offset of the rotational transform profiles
associated with each setting of $I_{IL}/I_{PF}$. Hence, incorporating CNT's
field errors is expected to result in an offset of the plots shown in 
Fig.~\ref{fig:maxBeta} along the abscissa.

\section{Heating power requirements}
\label{sect:scaling}

The heating power
requirements for the scenarios outlined in the previous
section can be roughly estimated as follows stellarator scaling laws.

We estimate $\beta$ as\cite{chen1986}
\begin{equation}
\label{eqn:beta}
    \beta  = \frac{n_ek_BT_e + n_ik_BT_i}{B^2/2\mu_0}. 
\end{equation}

\noindent Here, $n_e$ and $n_i$ are the electron and ion densities (assuming
a single dominant ion species), $T_e$ and $T_i$ are the electron and ion 
temperatures, $k_B$ is Boltzmann's constant, $B$ is the 
magnetic field, and $\mu_0$ is the vacuum permeability. We assume a 
quasi-neutral plasma with $n_e = n_i$ and, as in Sec.~\ref{sect:inputParams},
$T_i = 0.3T_e$. Hence, the numerator of Eq.~\ref{eqn:beta} simplifies to
$1.3n_ek_BT_e$. This, in turn, may be re-expressed in terms of heating power
$P$ as $1.3(\tau_EP/V)$, where $V$ is the plasma volume and 
where $\tau_E$ is the energy confinement time.

We estimate $\tau_E$ using the 2004 International Stellarator Scaling law 
(ISS04):\cite{yamada2005}

\begin{equation}
\label{eqn:tauE}
    \tau_\text{E,ISS04} = 0.134 f_\text{ren} 
             a^{2.28} R^{0.64} P^{-0.61} \bar{n}_e^{0.54} B^{0.84} 
             \text{\sout{$\iota$}}_{2/3}^{0.41}
\end{equation}

\noindent Here $R$ and $a$ are the plasma major and minor 
radius in meters, $B$ is the magnetic field in Tesla, and 
$\text{\sout{$\iota$}}_{2/3}$ is the rotational transform evaluated at 
two-thirds the minor radius of the LCFS. The heating
power $P$ (in MW in this formula) is treated as
an independent variable. Incidentally, the ISS04 dataset
involved several heliotrons/torsatrons and one device with circular coils
(the TJ-II heliac). Also note that CNT is essentially a heliotron/torsatron
(not a classical stellarator) with two ($\ell=2$) ``helical'' coils of poloidal 
number $m = 1$ and toroidal number $n = 1$ which, in effect, are circular.
It should be noted that the values of $R$, $B$, and
line-averaged electron density $\bar{n}_e$ (in units of 
$10^{19}~\text{m}^{-3}$), are lower in CNT than in other devices which 
$\tau_\text{E,ISS04}$ is based upon. Said otherwise, the scaling law is being
extrapolated here. Furthermore, $\tau_\text{E,ISS04}$ is determined based on 
current-free or nearly current-free plasmas, whereas the high-$\beta$, 
low-aspect-ratio equilibria evaluated in this paper contain a small
but finite bootstrap current. 

The renormalization coefficient, $f_\text{ren}$, is a device-specific fitting
parameter. Among the stellarators in the ISS04 database, 
this value ranged from 0.25 for TJ-II to unity for a 
configuration of W7-AS. This parameter has not been calculated for CNT. 
However, it was observed that the parameter was inversely correlated with the
toroidal effective ripple, $\epsilon_\text{eff}$. This 
tendency follows from the association of greater ripple with a larger
population of helically trapped particles. In particular, 
$\epsilon_\text{eff}(2/3)$ (evaulated at two-thirds of the minor radius) 
was found to relate to $f_\text{ren}$ roughly as 
$f_\text{ren} \propto \epsilon_\text{eff}(2/3)^{-0.4}$ for values of 
$\epsilon_\text{eff}(2/3)$ between 0.02 and 0.4 (see Fig.~7 in 
Ref.~\cite{yamada2005}).
For CNT, a value of $\epsilon_\text{eff}(2/3)$ of 1.6 was calculated for 
the configuration $\theta_\text{tilt}=64^\circ$, $I_{IL}/I_{PF}=4.22$.
\cite{seiwald2007}. Extrapolating the relationship observed in the ISS04
database to the range of $\epsilon_\text{eff}$ for CNT, we posit 
$f_\text{ren} \approx (0.25 \pm 0.05)\epsilon_\text{eff}(2/3)^{-0.4}$, leading
to an estimate of $f_\text{ren} = 0.21$ for CNT with upper and lower 
bounds of 0.25 and 0.17, respectively.
%a lower 
%limit of roughly 0.1 for $f_\text{ren}$. In the remainder, we adopt this lowest,
%most pessimistic value, $f_\text{ren}=0.1$, for a conservative estimate of
%$\tau_E$.

$\tau_E$ is evaluated in two main regimes of 
electron density density $\bar{n}_e$.
The first is the highest attainable before the plasma radiatively collapses, 
determined through a formula
derived by Sudo,\cite{sudo1990}

\begin{equation}
\label{eqn:nSudo}
    n_\text{Sudo} = 2.5\sqrt{\frac{PB}{a^2R}}.
\end{equation}

%\noindent or an empirical scaling law obtained from 
%W7-AS data,\cite{giannone2003} 

%\begin{equation}
%\label{eqn:nw7as}
%    n_\text{W7-AS} = 15\left(\frac{P}{V}\right)^{0.40}B^{0.32}
%\end{equation}

\noindent A similar formula was derived at W7-AS \cite{giannone2003} and yields
similar results, not shown for brevity. The second is the cutoff density above 
which microwaves for ECH 
cannot propagate. For heating at the fundamental harmonic in the ordinary
mode at frequency 
$\omega_\text{rf} = 2\pi\times2.45$ GHz, this is

\begin{equation}
\label{eqn:nco}
    n_\text{co,O} = \frac{\epsilon_0m_e}{e^2}{\omega_\text{rf}}^2.
\end{equation}

\noindent For heating at the second harmonic 
in the extraordinary mode using the same 
heating frequency (and, therefore, half the field), the highest 
density at which propagation may occur throughout the plasma 
corresponds to the right-handed cutoff, 
effectively half of the ordinary-mode cutoff:

\begin{equation}
\label{eqn:ncox2}
    n_\text{co,X2} = \frac{1}{2}\frac{\epsilon_0m_e}{e^2}{\omega_\text{rf}}^2;
\end{equation}

\begin{figure}
    \includegraphics[width=0.5\textwidth]{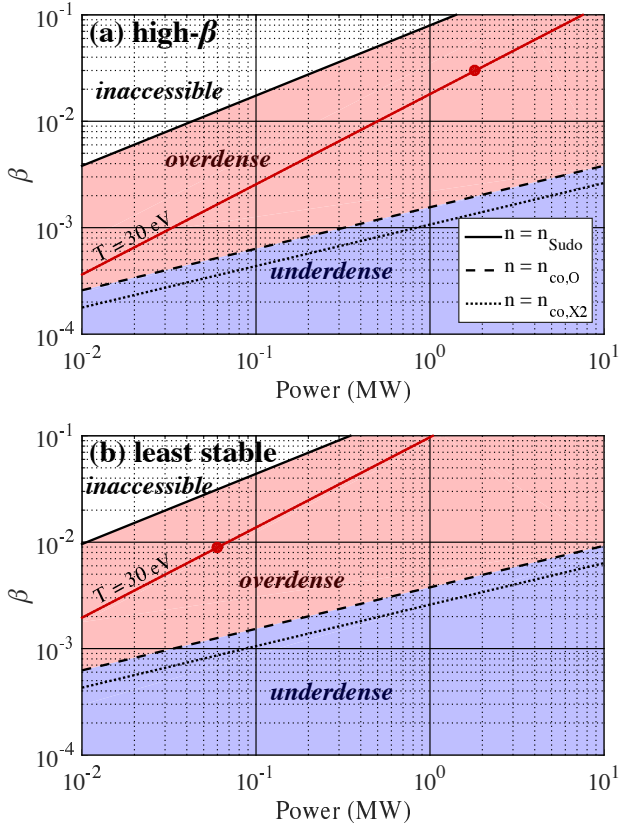}
    \caption{Accessible $\beta$ at different levels of heating power according
             to stellarator scaling laws for (a) the 
             high-$\beta$ configuration and (b) the least stable configuration.
             Values of $\beta$ above the solid line are considered inaccessible 
             because the plasma density exceeds the Sudo limit. Values below the
             dashed line correspond to plasmas underdense to 
             2.45 GHz ECH in the ordinary mode; values between the solid 
             and dashed lines could be attainable by overdense heating.
             The red lines are contours of $T_e = 30$ eV in the
             $(P,\beta)$ parameter space. The red circles indicate the power 
             levels needed to attain the maximum ballooning-stable $\beta$ in 
             each configuration.}
    \label{fig:scaling}
\end{figure}

Results of these calculations for the high-$\beta$ and least 
stable configurations
are shown in Fig.~\ref{fig:scaling}. The three black curves shown 
in each plot, from top to bottom, give the value of $\beta$ 
determined using Eqs.~\ref{eqn:beta}-\ref{eqn:tauE}, with densities from 
Eqs.~\ref{eqn:nSudo}, \ref{eqn:nco}, and \ref{eqn:ncox2}, respectively. 
%(Note that, for all values of 
%$P$, the Sudo and W7-AS density limits agree to within a factor $< 1.8$ and 
%the propagation cutoffs agree even more closely.) 
Thus, the blue-shaded region below the curve for $\beta(n_\text{co,O})$ is 
accessible by underdense plasmas, and the red-shaded region between 
$\beta(n_\text{co,O})$ and $\beta(n_\text{Sudo})$ corresponds to $\beta$ 
values that are accessible with overdense microwave heating. 

The red lines in Fig.~\ref{fig:scaling}a-b are $(P,\beta)$ 
contours
corresponding to $T_e = 30$ eV, determined by inverting Eqs.~\ref{eqn:beta} and 
\ref{eqn:tauE}. To the left of these lines, $T_e$ is lower; to the right, $T_e$
is higher. Thus, to obtain $\beta = 3.0\%$ in the high-$\beta$ configuration
while maintaining a minimum $T_e$ of 30 eV, about 1.8 MW of power will be 
needed 
(Fig.~\ref{fig:scaling}a). To obtain $\beta = 0.9\%$ in the least stable 
configuration at $T_e = 30$ eV, about 60 kW will be needed. It should be 
noted, however, that these estimates are sensitive to $f_\text{ren}$
(Eq.~\ref{eqn:tauE}). If these estimates are redone using the upper and lower
bounds mentioned above ($0.17 < f_\text{ren} < 0.25$), a range of 1 to 3 MW is 
established for the high-$\beta$ configuration and 40 to 100 kW for the 
least stable configuration. 

Both configurations fall within the overdense region for 
heating with 2.45 GHz and will therefore both require overdense microwave 
heating. The parameters of
these configurations are compared with present experimental conditions in 
Table~\ref{tab:param_compare}. Note that $\text{\sout{$\iota$}}_{2/3}$ for 
the least stable configuration exceeds that of the high-$\beta$ configuration
by a factor of nearly 7. This corresponds to a factor of 2 increase in 
$\tau_\text{E}$ (Eq.~\ref{eqn:tauE}) due to $\text{\sout{$\iota$}}_{2/3}$ 
alone. This, combined with the reduced $\beta$, explains why so much less power 
is required for the least stable configuration.

%Thus, the high-$\beta$
%configuration with $\beta = 3.7\%$ cannot be attained with $P<1$ MW if the 
%plasma remains underdense, but might be possible with heating powers as low 
%as $\approx$ 200 kW if the plasma may become overdense.

\begin{table}
    \begin{tabular}{| c c c c |}
        \hline
        ~ Parameter ~ & ~ Present ~ & ~ Least stable ~ & ~ High-$\beta$ ~ \\ 
        \hline
        $\theta_\text{tilt}$ (deg) & 78 & 88  & 78 \\ 
        $a$ (m) & 0.14  & 0.12 & 0.13 \\ 
        $R$ (m) & 0.31  & 0.30 & 0.32  \\ 
        $P$ (kW) & 0.5-8 & 40-100 & ~ 1,000-3,000 ~ \\ 
        ~ $n_e$ ($10^{17}~\text{m}^{-3}$) ~ &  0.5-3 & 50 & 200 \\
        $B$ (T) & 0.08 & 0.08 & 0.08 \\ 
        $\text{\sout{$\iota$}}_{2/3}$ & 0.36 & 0.62 & 0.09 \\ 
        $T_e$ (eV) & 4-8 & 30 & 30 \\ \hline
    \end{tabular}
\caption{Comparison of present CNT experimental parameters with those of the
         least stable and high-$\beta$ configurations.}
\label{tab:param_compare}
\end{table}

\section{Conclusions and future work}
\label{sect:vmecConclusion}

The foregoing work has established that ballooning-stable equilibria with 
average $\beta$ values of up to 3.0\% should be attainable in 
CNT, and possibly even higher, as the ballooning stability
limit was not found for some configurations (Fig.~\ref{fig:maxBeta}a). 
Here it was assumed that the plasma current is dominated by the bootstrap 
current. The limitations on stability depend on the magnetic configuration, 
with the $\theta_\text{tilt}=78^\circ$ coil orientation 
capable of the highest $\beta$.
Stability also depends on the pressure profile and the associated bootstrap
current profile: plasmas with peaked pressure profiles, for the same 
$\theta_\text{tilt}$ and $I_{IL}/I_{PF}$, may become unstable at 
$\beta\simeq1.4\%$. These stability limits are lower than in 
other stellarators, and thus more easily accessible.
Stellarator scaling laws indicate that a 30 eV 
plasma with $\beta=3.0\%$ might be attainable with 2.45 GHz microwaves 
at a power of 1-3 MW if overdense heating mechanisms are
employed. Furthermore, a configuration was found that may
become ballooning-unstable at $\beta$ as low as 0.9\%, which would be 
attainable at a power of 40-100 kW.

%The equilibrium parameters and bootstrap current profile 
%were obtained using experimental temperature and 
%pressure profiles from overdense CNT plasmas heated with less than 10 kW.
A self-consistent estimate of the maximum stable $\beta$ would require that,
instead of the normalized profiles in Fig.~\ref{fig:presProf} 
based on 10 kW heating experiments, we use for the 
stability calculations a pressure profile as close as possible to what we would
obtain with higher heating power. In turn,
this would require full-wave and Fokker-Planck modeling, which is left as
future work. A ray- or beam-tracing would not
be appropriate due to the spatial 
scales of the problem (\textit{i.e.}, the plasma minor
radius is similar to the microwave vacuum wavelength).

Finally, indications of a second region of stability were 
found in Figs.~\ref{fig:qty_vs_beta}-\ref{fig:stab_contours}, which will 
deserve more extensive modeling and experimentation in the future.


\begin{thebibliography}{31}%
\makeatletter
\providecommand \@ifxundefined [1]{%
 \@ifx{#1\undefined}
}%
\providecommand \@ifnum [1]{%
 \ifnum #1\expandafter \@firstoftwo
 \else \expandafter \@secondoftwo
 \fi
}%
\providecommand \@ifx [1]{%
 \ifx #1\expandafter \@firstoftwo
 \else \expandafter \@secondoftwo
 \fi
}%
\providecommand \natexlab [1]{#1}%
\providecommand \enquote  [1]{``#1''}%
\providecommand \bibnamefont  [1]{#1}%
\providecommand \bibfnamefont [1]{#1}%
\providecommand \citenamefont [1]{#1}%
\providecommand \href@noop [0]{\@secondoftwo}%
\providecommand \href [0]{\begingroup \@sanitize@url \@href}%
\providecommand \@href[1]{\@@startlink{#1}\@@href}%
\providecommand \@@href[1]{\endgroup#1\@@endlink}%
\providecommand \@sanitize@url [0]{\catcode `\\12\catcode `\$12\catcode
  `\&12\catcode `\#12\catcode `\^12\catcode `\_12\catcode `\%12\relax}%
\providecommand \@@startlink[1]{}%
\providecommand \@@endlink[0]{}%
\providecommand \url  [0]{\begingroup\@sanitize@url \@url }%
\providecommand \@url [1]{\endgroup\@href {#1}{\urlprefix }}%
\providecommand \urlprefix  [0]{URL }%
\providecommand \Eprint [0]{\href }%
\providecommand \doibase [0]{http://dx.doi.org/}%
\providecommand \selectlanguage [0]{\@gobble}%
\providecommand \bibinfo  [0]{\@secondoftwo}%
\providecommand \bibfield  [0]{\@secondoftwo}%
\providecommand \translation [1]{[#1]}%
\providecommand \BibitemOpen [0]{}%
\providecommand \bibitemStop [0]{}%
\providecommand \bibitemNoStop [0]{.\EOS\space}%
\providecommand \EOS [0]{\spacefactor3000\relax}%
\providecommand \BibitemShut  [1]{\csname bibitem#1\endcsname}%
\let\auto@bib@innerbib\@empty
%</preamble>
\bibitem [{\citenamefont {Weller}\ \emph {et~al.}(2003)\citenamefont {Weller},
  \citenamefont {Geiger}, \citenamefont {Werner}, \citenamefont {Zarnstorff},
  \citenamefont {N\"{u}hrenberg}, \citenamefont {Sallander}, \citenamefont
  {Baldzuhn}, \citenamefont {Brakel}, \citenamefont {Burhenn}, \citenamefont
  {Dinklage}, \citenamefont {Fredrickson}, \citenamefont {Gadelmeier},
  \citenamefont {Gianonne}, \citenamefont {Grigull}, \citenamefont {Hartmann},
  \citenamefont {Jaenicke}, \citenamefont {Klose}, \citenamefont {Knauer},
  \citenamefont {K\"{o}nies}, \citenamefont {Kolesnichenko}, \citenamefont
  {Laqua}, \citenamefont {Lutsenko}, \citenamefont {McCormick}, \citenamefont
  {Monticello}, \citenamefont {Osakakabe}, \citenamefont {Pasch}, \citenamefont
  {Reiman}, \citenamefont {Rust}, \citenamefont {Spong}, \citenamefont
  {Wagner}, \citenamefont {Yakovenko}, \citenamefont {the W7-AS~Team},\ and\
  \citenamefont {NBI-Group}}]{weller2003}%
  \BibitemOpen
  \bibfield  {author} {\bibinfo {author} {\bibfnamefont {A.}~\bibnamefont
  {Weller}}, \bibinfo {author} {\bibfnamefont {J.}~\bibnamefont {Geiger}},
  \bibinfo {author} {\bibfnamefont {A.}~\bibnamefont {Werner}}, \bibinfo
  {author} {\bibfnamefont {M.~C.}\ \bibnamefont {Zarnstorff}}, \bibinfo
  {author} {\bibfnamefont {C.}~\bibnamefont {N\"{u}hrenberg}}, \bibinfo
  {author} {\bibfnamefont {E.}~\bibnamefont {Sallander}}, \bibinfo {author}
  {\bibfnamefont {J.}~\bibnamefont {Baldzuhn}}, \bibinfo {author}
  {\bibfnamefont {R.}~\bibnamefont {Brakel}}, \bibinfo {author} {\bibfnamefont
  {R.}~\bibnamefont {Burhenn}}, \bibinfo {author} {\bibfnamefont
  {A.}~\bibnamefont {Dinklage}}, \bibinfo {author} {\bibfnamefont
  {E.}~\bibnamefont {Fredrickson}}, \bibinfo {author} {\bibfnamefont
  {F.}~\bibnamefont {Gadelmeier}}, \bibinfo {author} {\bibfnamefont
  {L.}~\bibnamefont {Gianonne}}, \bibinfo {author} {\bibfnamefont
  {P.}~\bibnamefont {Grigull}}, \bibinfo {author} {\bibfnamefont
  {D.}~\bibnamefont {Hartmann}}, \bibinfo {author} {\bibfnamefont
  {R.}~\bibnamefont {Jaenicke}}, \bibinfo {author} {\bibfnamefont
  {S.}~\bibnamefont {Klose}}, \bibinfo {author} {\bibfnamefont {J.~P.}\
  \bibnamefont {Knauer}}, \bibinfo {author} {\bibfnamefont {A.}~\bibnamefont
  {K\"{o}nies}}, \bibinfo {author} {\bibfnamefont {Y.~I.}\ \bibnamefont
  {Kolesnichenko}}, \bibinfo {author} {\bibfnamefont {H.~P.}\ \bibnamefont
  {Laqua}}, \bibinfo {author} {\bibfnamefont {V.~V.}\ \bibnamefont {Lutsenko}},
  \bibinfo {author} {\bibfnamefont {K.}~\bibnamefont {McCormick}}, \bibinfo
  {author} {\bibfnamefont {D.}~\bibnamefont {Monticello}}, \bibinfo {author}
  {\bibfnamefont {M.}~\bibnamefont {Osakakabe}}, \bibinfo {author}
  {\bibfnamefont {E.}~\bibnamefont {Pasch}}, \bibinfo {author} {\bibfnamefont
  {A.}~\bibnamefont {Reiman}}, \bibinfo {author} {\bibfnamefont
  {N.}~\bibnamefont {Rust}}, \bibinfo {author} {\bibfnamefont {D.~A.}\
  \bibnamefont {Spong}}, \bibinfo {author} {\bibfnamefont {F.}~\bibnamefont
  {Wagner}}, \bibinfo {author} {\bibfnamefont {Y.~V.}\ \bibnamefont
  {Yakovenko}}, \bibinfo {author} {\bibnamefont {the W7-AS~Team}}, \ and\
  \bibinfo {author} {\bibnamefont {NBI-Group}},\ }\href@noop {} {\bibfield
  {journal} {\bibinfo  {journal} {Plasma Physics and Controlled Fusion}\
  }\textbf {\bibinfo {volume} {45}},\ \bibinfo {pages} {A285} (\bibinfo {year}
  {2003})}\BibitemShut {NoStop}%
\bibitem [{\citenamefont {Sakakibara}\ \emph {et~al.}(2015)\citenamefont
  {Sakakibara}, \citenamefont {Watanabe}, \citenamefont {Takemura},
  \citenamefont {Okamoto}, \citenamefont {Ohdachi}, \citenamefont {Suzuki},
  \citenamefont {Narushima}, \citenamefont {Ida}, \citenamefont {Yoshinuma},
  \citenamefont {Tanaka}, \citenamefont {Tokuzawa}, \citenamefont {Yamada},
  \citenamefont {Yamada}, \citenamefont {Takeiri},\ and\ \citenamefont {the LHD
  Experiment~Group}}]{sakakibara2015}%
  \BibitemOpen
  \bibfield  {author} {\bibinfo {author} {\bibfnamefont {S.}~\bibnamefont
  {Sakakibara}}, \bibinfo {author} {\bibfnamefont {K.~Y.}\ \bibnamefont
  {Watanabe}}, \bibinfo {author} {\bibfnamefont {Y.}~\bibnamefont {Takemura}},
  \bibinfo {author} {\bibfnamefont {M.}~\bibnamefont {Okamoto}}, \bibinfo
  {author} {\bibfnamefont {S.}~\bibnamefont {Ohdachi}}, \bibinfo {author}
  {\bibfnamefont {Y.}~\bibnamefont {Suzuki}}, \bibinfo {author} {\bibfnamefont
  {Y.}~\bibnamefont {Narushima}}, \bibinfo {author} {\bibfnamefont
  {K.}~\bibnamefont {Ida}}, \bibinfo {author} {\bibfnamefont {M.}~\bibnamefont
  {Yoshinuma}}, \bibinfo {author} {\bibfnamefont {K.}~\bibnamefont {Tanaka}},
  \bibinfo {author} {\bibfnamefont {T.}~\bibnamefont {Tokuzawa}}, \bibinfo
  {author} {\bibfnamefont {I.}~\bibnamefont {Yamada}}, \bibinfo {author}
  {\bibfnamefont {H.}~\bibnamefont {Yamada}}, \bibinfo {author} {\bibfnamefont
  {Y.}~\bibnamefont {Takeiri}}, \ and\ \bibinfo {author} {\bibnamefont {the LHD
  Experiment~Group}},\ }\href@noop {} {\bibfield  {journal} {\bibinfo
  {journal} {Nuclear Fusion}\ }\textbf {\bibinfo {volume} {55}},\ \bibinfo
  {pages} {083020} (\bibinfo {year} {2015})}\BibitemShut {NoStop}%
\bibitem [{\citenamefont {Cuthbert}\ \emph {et~al.}(1998)\citenamefont
  {Cuthbert}, \citenamefont {Lewandowski}, \citenamefont {Gardner},
  \citenamefont {Persson}, \citenamefont {Singleton}, \citenamefont {Dewar},
  \citenamefont {Nakajima},\ and\ \citenamefont {Cooper}}]{cuthbert1998}%
  \BibitemOpen
  \bibfield  {author} {\bibinfo {author} {\bibfnamefont {P.}~\bibnamefont
  {Cuthbert}}, \bibinfo {author} {\bibfnamefont {J.~L.~V.}\ \bibnamefont
  {Lewandowski}}, \bibinfo {author} {\bibfnamefont {H.~J.}\ \bibnamefont
  {Gardner}}, \bibinfo {author} {\bibfnamefont {M.}~\bibnamefont {Persson}},
  \bibinfo {author} {\bibfnamefont {D.~B.}\ \bibnamefont {Singleton}}, \bibinfo
  {author} {\bibfnamefont {R.~L.}\ \bibnamefont {Dewar}}, \bibinfo {author}
  {\bibfnamefont {N.}~\bibnamefont {Nakajima}}, \ and\ \bibinfo {author}
  {\bibfnamefont {W.~A.}\ \bibnamefont {Cooper}},\ }\href@noop {} {\bibfield
  {journal} {\bibinfo  {journal} {Physics of Plasmas}\ }\textbf {\bibinfo
  {volume} {5}},\ \bibinfo {pages} {2921} (\bibinfo {year} {1998})}\BibitemShut
  {NoStop}%
\bibitem [{\citenamefont {Hegna}\ and\ \citenamefont
  {Nakajima}(1998)}]{hegna1998}%
  \BibitemOpen
  \bibfield  {author} {\bibinfo {author} {\bibfnamefont {C.~C.}\ \bibnamefont
  {Hegna}}\ and\ \bibinfo {author} {\bibfnamefont {N.}~\bibnamefont
  {Nakajima}},\ }\href@noop {} {\bibfield  {journal} {\bibinfo  {journal}
  {Physics of Plasmas}\ }\textbf {\bibinfo {volume} {5}},\ \bibinfo {pages}
  {1336} (\bibinfo {year} {1998})}\BibitemShut {NoStop}%
\bibitem [{\citenamefont {Hudson}\ and\ \citenamefont
  {Hegna}(2003)}]{hudson2003}%
  \BibitemOpen
  \bibfield  {author} {\bibinfo {author} {\bibfnamefont {S.~R.}\ \bibnamefont
  {Hudson}}\ and\ \bibinfo {author} {\bibfnamefont {C.~C.}\ \bibnamefont
  {Hegna}},\ }\href@noop {} {\bibfield  {journal} {\bibinfo  {journal} {Physics
  of Plasmas}\ }\textbf {\bibinfo {volume} {10}},\ \bibinfo {pages} {4716}
  (\bibinfo {year} {2003})}\BibitemShut {NoStop}%
\bibitem [{\citenamefont {Hudson}\ and\ \citenamefont
  {Hegna}(2004)}]{hudson2004}%
  \BibitemOpen
  \bibfield  {author} {\bibinfo {author} {\bibfnamefont {S.~R.}\ \bibnamefont
  {Hudson}}\ and\ \bibinfo {author} {\bibfnamefont {C.~C.}\ \bibnamefont
  {Hegna}},\ }\href@noop {} {\bibfield  {journal} {\bibinfo  {journal} {Physics
  of Plasmas}\ }\textbf {\bibinfo {volume} {11}},\ \bibinfo {pages} {L53}
  (\bibinfo {year} {2004})}\BibitemShut {NoStop}%
\bibitem [{\citenamefont {Rafiq}\ \emph {et~al.}(2010)\citenamefont {Rafiq},
  \citenamefont {Hegna}, \citenamefont {Callen},\ and\ \citenamefont
  {Kritz}}]{rafiq2010}%
  \BibitemOpen
  \bibfield  {author} {\bibinfo {author} {\bibfnamefont {T.}~\bibnamefont
  {Rafiq}}, \bibinfo {author} {\bibfnamefont {C.~C.}\ \bibnamefont {Hegna}},
  \bibinfo {author} {\bibfnamefont {J.~D.}\ \bibnamefont {Callen}}, \ and\
  \bibinfo {author} {\bibfnamefont {A.~H.}\ \bibnamefont {Kritz}},\ }\href@noop
  {} {\bibfield  {journal} {\bibinfo  {journal} {Physics of Plasmas}\ }\textbf
  {\bibinfo {volume} {17}},\ \bibinfo {pages} {022502} (\bibinfo {year}
  {2010})}\BibitemShut {NoStop}%
\bibitem [{\citenamefont {Nielson}\ \emph {et~al.}(2000)\citenamefont
  {Nielson}, \citenamefont {Reiman}, \citenamefont {Zarnstorff}, \citenamefont
  {Brooks}, \citenamefont {Fu}, \citenamefont {Goldston}, \citenamefont {Ku},
  \citenamefont {Lin}, \citenamefont {Majeski}, \citenamefont {Monticello},
  \citenamefont {Mynick}, \citenamefont {Pomphrey}, \citenamefont {Redi},
  \citenamefont {Reierson}, \citenamefont {Schmidt}, \citenamefont {Hirshman},
  \citenamefont {Lyon}, \citenamefont {Berry}, \citenamefont {Nelson},
  \citenamefont {Sanchez}, \citenamefont {Spong}, \citenamefont {Boozer},
  \citenamefont {{Miner Jr.}}, \citenamefont {Valanju}, \citenamefont {Cooper},
  \citenamefont {Drevlak}, \citenamefont {Merkel},\ and\ \citenamefont
  {N{\"u}hrenberg}}]{nielson2000}%
  \BibitemOpen
  \bibfield  {author} {\bibinfo {author} {\bibfnamefont {G.~H.}\ \bibnamefont
  {Nielson}}, \bibinfo {author} {\bibfnamefont {A.~H.}\ \bibnamefont {Reiman}},
  \bibinfo {author} {\bibfnamefont {M.~C.}\ \bibnamefont {Zarnstorff}},
  \bibinfo {author} {\bibfnamefont {A.}~\bibnamefont {Brooks}}, \bibinfo
  {author} {\bibfnamefont {G.-Y.}\ \bibnamefont {Fu}}, \bibinfo {author}
  {\bibfnamefont {R.~J.}\ \bibnamefont {Goldston}}, \bibinfo {author}
  {\bibfnamefont {L.-P.}\ \bibnamefont {Ku}}, \bibinfo {author} {\bibfnamefont
  {Z.}~\bibnamefont {Lin}}, \bibinfo {author} {\bibfnamefont {R.}~\bibnamefont
  {Majeski}}, \bibinfo {author} {\bibfnamefont {D.~A.}\ \bibnamefont
  {Monticello}}, \bibinfo {author} {\bibfnamefont {H.}~\bibnamefont {Mynick}},
  \bibinfo {author} {\bibfnamefont {N.}~\bibnamefont {Pomphrey}}, \bibinfo
  {author} {\bibfnamefont {M.~H.}\ \bibnamefont {Redi}}, \bibinfo {author}
  {\bibfnamefont {W.~T.}\ \bibnamefont {Reierson}}, \bibinfo {author}
  {\bibfnamefont {J.~A.}\ \bibnamefont {Schmidt}}, \bibinfo {author}
  {\bibfnamefont {S.~P.}\ \bibnamefont {Hirshman}}, \bibinfo {author}
  {\bibfnamefont {J.~F.}\ \bibnamefont {Lyon}}, \bibinfo {author}
  {\bibfnamefont {L.~A.}\ \bibnamefont {Berry}}, \bibinfo {author}
  {\bibfnamefont {B.~E.}\ \bibnamefont {Nelson}}, \bibinfo {author}
  {\bibfnamefont {R.}~\bibnamefont {Sanchez}}, \bibinfo {author} {\bibfnamefont
  {D.~A.}\ \bibnamefont {Spong}}, \bibinfo {author} {\bibfnamefont {A.~H.}\
  \bibnamefont {Boozer}}, \bibinfo {author} {\bibfnamefont {W.~H.}\
  \bibnamefont {{Miner Jr.}}}, \bibinfo {author} {\bibfnamefont {P.~M.}\
  \bibnamefont {Valanju}}, \bibinfo {author} {\bibfnamefont {W.~A.}\
  \bibnamefont {Cooper}}, \bibinfo {author} {\bibfnamefont {M.}~\bibnamefont
  {Drevlak}}, \bibinfo {author} {\bibfnamefont {P.}~\bibnamefont {Merkel}}, \
  and\ \bibinfo {author} {\bibfnamefont {C.}~\bibnamefont {N{\"u}hrenberg}},\
  }\href@noop {} {\bibfield  {journal} {\bibinfo  {journal} {Physics of
  Plasmas}\ }\textbf {\bibinfo {volume} {7}},\ \bibinfo {pages} {1911}
  (\bibinfo {year} {2000})}\BibitemShut {NoStop}%
\bibitem [{\citenamefont {Weller}\ \emph {et~al.}(2001)\citenamefont {Weller},
  \citenamefont {Anton}, \citenamefont {Geiger}, \citenamefont {Hirsch},
  \citenamefont {Jaenicke}, \citenamefont {Werner}, \citenamefont
  {N{\"u}hrenberg}, \citenamefont {Sallander}, \citenamefont {Spong},\ and\
  \citenamefont {the W7-AS~team}}]{weller2001}%
  \BibitemOpen
  \bibfield  {author} {\bibinfo {author} {\bibfnamefont {A.}~\bibnamefont
  {Weller}}, \bibinfo {author} {\bibfnamefont {M.}~\bibnamefont {Anton}},
  \bibinfo {author} {\bibfnamefont {J.}~\bibnamefont {Geiger}}, \bibinfo
  {author} {\bibfnamefont {M.}~\bibnamefont {Hirsch}}, \bibinfo {author}
  {\bibfnamefont {R.}~\bibnamefont {Jaenicke}}, \bibinfo {author}
  {\bibfnamefont {A.}~\bibnamefont {Werner}}, \bibinfo {author} {\bibfnamefont
  {C.}~\bibnamefont {N{\"u}hrenberg}}, \bibinfo {author} {\bibfnamefont
  {E.}~\bibnamefont {Sallander}}, \bibinfo {author} {\bibfnamefont {D.~A.}\
  \bibnamefont {Spong}}, \ and\ \bibinfo {author} {\bibnamefont {the
  W7-AS~team}},\ }\href@noop {} {\bibfield  {journal} {\bibinfo  {journal}
  {Physics of Plasmas}\ }\textbf {\bibinfo {volume} {8}},\ \bibinfo {pages}
  {931} (\bibinfo {year} {2001})}\BibitemShut {NoStop}%
\bibitem [{\citenamefont {Beidler}\ \emph {et~al.}(2001)\citenamefont
  {Beidler}, \citenamefont {Harmeyer}, \citenamefont {Herrnegger},
  \citenamefont {Igitkhanov}, \citenamefont {Kendl}, \citenamefont
  {Kisslinger}, \citenamefont {Kolesnichenko}, \citenamefont {Lutsenko},
  \citenamefont {N{\"u}hrenberg}, \citenamefont {Sidorenko}, \citenamefont
  {Strumberger}, \citenamefont {Wobig},\ and\ \citenamefont
  {Yakovenko}}]{beidler2001}%
  \BibitemOpen
  \bibfield  {author} {\bibinfo {author} {\bibfnamefont {C.~D.}\ \bibnamefont
  {Beidler}}, \bibinfo {author} {\bibfnamefont {E.}~\bibnamefont {Harmeyer}},
  \bibinfo {author} {\bibfnamefont {F.}~\bibnamefont {Herrnegger}}, \bibinfo
  {author} {\bibfnamefont {Y.}~\bibnamefont {Igitkhanov}}, \bibinfo {author}
  {\bibfnamefont {A.}~\bibnamefont {Kendl}}, \bibinfo {author} {\bibfnamefont
  {J.}~\bibnamefont {Kisslinger}}, \bibinfo {author} {\bibfnamefont {Y.~I.}\
  \bibnamefont {Kolesnichenko}}, \bibinfo {author} {\bibfnamefont {V.~V.}\
  \bibnamefont {Lutsenko}}, \bibinfo {author} {\bibfnamefont {C.}~\bibnamefont
  {N{\"u}hrenberg}}, \bibinfo {author} {\bibfnamefont {I.}~\bibnamefont
  {Sidorenko}}, \bibinfo {author} {\bibfnamefont {E.}~\bibnamefont
  {Strumberger}}, \bibinfo {author} {\bibfnamefont {H.}~\bibnamefont {Wobig}},
  \ and\ \bibinfo {author} {\bibfnamefont {Y.~V.}\ \bibnamefont {Yakovenko}},\
  }\href@noop {} {\bibfield  {journal} {\bibinfo  {journal} {Nuclear Fusion}\
  }\textbf {\bibinfo {volume} {41}},\ \bibinfo {pages} {1759} (\bibinfo {year}
  {2001})}\BibitemShut {NoStop}%
\bibitem [{\citenamefont {Sagara}, \citenamefont {Igitkhanov},\ and\
  \citenamefont {Najmabadi}(2010)}]{sagara2010}%
  \BibitemOpen
  \bibfield  {author} {\bibinfo {author} {\bibfnamefont {A.}~\bibnamefont
  {Sagara}}, \bibinfo {author} {\bibfnamefont {Y.}~\bibnamefont {Igitkhanov}},
  \ and\ \bibinfo {author} {\bibfnamefont {F.}~\bibnamefont {Najmabadi}},\
  }\href@noop {} {\bibfield  {journal} {\bibinfo  {journal} {Fusion Engineering
  and Design}\ }\textbf {\bibinfo {volume} {85}},\ \bibinfo {pages} {1336}
  (\bibinfo {year} {2010})}\BibitemShut {NoStop}%
\bibitem [{\citenamefont {Pedersen}\ \emph {et~al.}(2004)\citenamefont
  {Pedersen}, \citenamefont {Boozer}, \citenamefont {Kremer}, \citenamefont
  {Lefrancois}, \citenamefont {Reiersen}, \citenamefont {Dahlgren},\ and\
  \citenamefont {Pomphrey}}]{pedersen2004}%
  \BibitemOpen
  \bibfield  {author} {\bibinfo {author} {\bibfnamefont {T.~S.}\ \bibnamefont
  {Pedersen}}, \bibinfo {author} {\bibfnamefont {A.~H.}\ \bibnamefont
  {Boozer}}, \bibinfo {author} {\bibfnamefont {J.~P.}\ \bibnamefont {Kremer}},
  \bibinfo {author} {\bibfnamefont {R.~G.}\ \bibnamefont {Lefrancois}},
  \bibinfo {author} {\bibfnamefont {W.~T.}\ \bibnamefont {Reiersen}}, \bibinfo
  {author} {\bibfnamefont {F.~D.}\ \bibnamefont {Dahlgren}}, \ and\ \bibinfo
  {author} {\bibfnamefont {N.}~\bibnamefont {Pomphrey}},\ }\href@noop {}
  {\bibfield  {journal} {\bibinfo  {journal} {Fusion Sci. Technol.}\ }\textbf
  {\bibinfo {volume} {46}},\ \bibinfo {pages} {200} (\bibinfo {year}
  {2004})}\BibitemShut {NoStop}%
\bibitem [{\citenamefont {Pedersen}\ \emph {et~al.}(2006)\citenamefont
  {Pedersen}, \citenamefont {Kremer}, \citenamefont {Lefrancois}, \citenamefont
  {Marksteiner}, \citenamefont {Sarasola},\ and\ \citenamefont
  {Ahmad}}]{pedersen_pop2006}%
  \BibitemOpen
  \bibfield  {author} {\bibinfo {author} {\bibfnamefont {T.~S.}\ \bibnamefont
  {Pedersen}}, \bibinfo {author} {\bibfnamefont {J.~P.}\ \bibnamefont
  {Kremer}}, \bibinfo {author} {\bibfnamefont {R.~G.}\ \bibnamefont
  {Lefrancois}}, \bibinfo {author} {\bibfnamefont {Q.}~\bibnamefont
  {Marksteiner}}, \bibinfo {author} {\bibfnamefont {X.}~\bibnamefont
  {Sarasola}}, \ and\ \bibinfo {author} {\bibfnamefont {N.}~\bibnamefont
  {Ahmad}},\ }\href@noop {} {\bibfield  {journal} {\bibinfo  {journal} {Phys.
  Plasmas}\ }\textbf {\bibinfo {volume} {13}},\ \bibinfo {pages} {012502}
  (\bibinfo {year} {2006})}\BibitemShut {NoStop}%
\bibitem [{\citenamefont {Kremer}\ \emph {et~al.}(2006)\citenamefont {Kremer},
  \citenamefont {Pedersen}, \citenamefont {Lefrancois},\ and\ \citenamefont
  {Marksteiner}}]{kremer2006}%
  \BibitemOpen
  \bibfield  {author} {\bibinfo {author} {\bibfnamefont {J.~P.}\ \bibnamefont
  {Kremer}}, \bibinfo {author} {\bibfnamefont {T.~S.}\ \bibnamefont
  {Pedersen}}, \bibinfo {author} {\bibfnamefont {R.~G.}\ \bibnamefont
  {Lefrancois}}, \ and\ \bibinfo {author} {\bibfnamefont {Q.}~\bibnamefont
  {Marksteiner}},\ }\href@noop {} {\bibfield  {journal} {\bibinfo  {journal}
  {Phys. Rev. Lett.}\ }\textbf {\bibinfo {volume} {97}},\ \bibinfo {pages}
  {095003} (\bibinfo {year} {2006})}\BibitemShut {NoStop}%
\bibitem [{\citenamefont {Sarasola}\ and\ \citenamefont
  {Pedersen}(2012)}]{sarasola2012}%
  \BibitemOpen
  \bibfield  {author} {\bibinfo {author} {\bibfnamefont {X.}~\bibnamefont
  {Sarasola}}\ and\ \bibinfo {author} {\bibfnamefont {T.~S.}\ \bibnamefont
  {Pedersen}},\ }\href@noop {} {\bibfield  {journal} {\bibinfo  {journal}
  {Plasma Phys. Controlled Fusion}\ }\textbf {\bibinfo {volume} {54}},\
  \bibinfo {pages} {124008} (\bibinfo {year} {2012})}\BibitemShut {NoStop}%
\bibitem [{\citenamefont {Hammond}\ \emph
  {et~al.}(2016{\natexlab{a}})\citenamefont {Hammond}, \citenamefont
  {Anichowski}, \citenamefont {Brenner}, \citenamefont {Pedersen},
  \citenamefont {Raftopoulos}, \citenamefont {Traverso},\ and\ \citenamefont
  {Volpe}}]{hammond2016}%
  \BibitemOpen
  \bibfield  {author} {\bibinfo {author} {\bibfnamefont {K.~C.}\ \bibnamefont
  {Hammond}}, \bibinfo {author} {\bibfnamefont {A.}~\bibnamefont {Anichowski}},
  \bibinfo {author} {\bibfnamefont {P.~W.}\ \bibnamefont {Brenner}}, \bibinfo
  {author} {\bibfnamefont {T.~S.}\ \bibnamefont {Pedersen}}, \bibinfo {author}
  {\bibfnamefont {S.}~\bibnamefont {Raftopoulos}}, \bibinfo {author}
  {\bibfnamefont {P.}~\bibnamefont {Traverso}}, \ and\ \bibinfo {author}
  {\bibfnamefont {F.~A.}\ \bibnamefont {Volpe}},\ }\href@noop {} {\bibfield
  {journal} {\bibinfo  {journal} {Plasma Physics and Controlled Fusion}\
  }\textbf {\bibinfo {volume} {58}},\ \bibinfo {pages} {074002} (\bibinfo
  {year} {2016}{\natexlab{a}})}\BibitemShut {NoStop}%
\bibitem [{\citenamefont {Hammond}\ \emph
  {et~al.}(2016{\natexlab{b}})\citenamefont {Hammond}, \citenamefont
  {Diaz-Pacheco}, \citenamefont {Kornbluth}, \citenamefont {Volpe},\ and\
  \citenamefont {Wei}}]{hammond_rsi2016}%
  \BibitemOpen
  \bibfield  {author} {\bibinfo {author} {\bibfnamefont {K.~C.}\ \bibnamefont
  {Hammond}}, \bibinfo {author} {\bibfnamefont {R.~R.}\ \bibnamefont
  {Diaz-Pacheco}}, \bibinfo {author} {\bibfnamefont {Y.}~\bibnamefont
  {Kornbluth}}, \bibinfo {author} {\bibfnamefont {F.~A.}\ \bibnamefont
  {Volpe}}, \ and\ \bibinfo {author} {\bibfnamefont {Y.}~\bibnamefont {Wei}},\
  }\href@noop {} {\bibfield  {journal} {\bibinfo  {journal} {Review of
  Scientific Instruments}\ }\textbf {\bibinfo {volume} {87}},\ \bibinfo {pages}
  {11E119} (\bibinfo {year} {2016}{\natexlab{b}})}\BibitemShut {NoStop}%
\bibitem [{\citenamefont {Hirshman}\ and\ \citenamefont
  {Whitson}(1983)}]{hirshman1983}%
  \BibitemOpen
  \bibfield  {author} {\bibinfo {author} {\bibfnamefont {S.~P.}\ \bibnamefont
  {Hirshman}}\ and\ \bibinfo {author} {\bibfnamefont {J.~C.}\ \bibnamefont
  {Whitson}},\ }\href@noop {} {\bibfield  {journal} {\bibinfo  {journal}
  {Physics of Fluids}\ }\textbf {\bibinfo {volume} {26}},\ \bibinfo {pages}
  {3553} (\bibinfo {year} {1983})}\BibitemShut {NoStop}%
\bibitem [{\citenamefont {Sanchez}\ \emph {et~al.}(2000)\citenamefont
  {Sanchez}, \citenamefont {Hirshman}, \citenamefont {Whitson},\ and\
  \citenamefont {Ware}}]{sanchez2000}%
  \BibitemOpen
  \bibfield  {author} {\bibinfo {author} {\bibfnamefont {R.}~\bibnamefont
  {Sanchez}}, \bibinfo {author} {\bibfnamefont {S.}~\bibnamefont {Hirshman}},
  \bibinfo {author} {\bibfnamefont {J.}~\bibnamefont {Whitson}}, \ and\
  \bibinfo {author} {\bibfnamefont {A.}~\bibnamefont {Ware}},\ }\href@noop {}
  {\bibfield  {journal} {\bibinfo  {journal} {Journal of Computational
  Physics}\ }\textbf {\bibinfo {volume} {161}},\ \bibinfo {pages} {576}
  (\bibinfo {year} {2000})}\BibitemShut {NoStop}%
\bibitem [{\citenamefont {Hirshman}, \citenamefont {van Rij},\ and\
  \citenamefont {Merkel}(1986)}]{hirshman1986}%
  \BibitemOpen
  \bibfield  {author} {\bibinfo {author} {\bibfnamefont {S.~P.}\ \bibnamefont
  {Hirshman}}, \bibinfo {author} {\bibfnamefont {W.~I.}\ \bibnamefont {van
  Rij}}, \ and\ \bibinfo {author} {\bibfnamefont {P.}~\bibnamefont {Merkel}},\
  }\href@noop {} {\bibfield  {journal} {\bibinfo  {journal} {Computer Physics
  Communications}\ }\textbf {\bibinfo {volume} {43}},\ \bibinfo {pages} {143}
  (\bibinfo {year} {1986})}\BibitemShut {NoStop}%
\bibitem [{\citenamefont {Weller}\ \emph {et~al.}(2006)\citenamefont {Weller},
  \citenamefont {Sakakibara}, \citenamefont {Watanabe}, \citenamefont {Toi},
  \citenamefont {Geiger}, \citenamefont {Zarnstorff}, \citenamefont {Hudson},
  \citenamefont {Reiman}, \citenamefont {Werner}, \citenamefont
  {N{\"u}hrenberg}, \citenamefont {Ohdachi}, \citenamefont {Suzuki},
  \citenamefont {Yamada}, \citenamefont {the W7-AS~team},\ and\ \citenamefont
  {the LHD~team}}]{weller2006}%
  \BibitemOpen
  \bibfield  {author} {\bibinfo {author} {\bibfnamefont {A.}~\bibnamefont
  {Weller}}, \bibinfo {author} {\bibfnamefont {S.}~\bibnamefont {Sakakibara}},
  \bibinfo {author} {\bibfnamefont {K.~Y.}\ \bibnamefont {Watanabe}}, \bibinfo
  {author} {\bibfnamefont {K.}~\bibnamefont {Toi}}, \bibinfo {author}
  {\bibfnamefont {J.}~\bibnamefont {Geiger}}, \bibinfo {author} {\bibfnamefont
  {M.~C.}\ \bibnamefont {Zarnstorff}}, \bibinfo {author} {\bibfnamefont
  {S.~R.}\ \bibnamefont {Hudson}}, \bibinfo {author} {\bibfnamefont
  {A.}~\bibnamefont {Reiman}}, \bibinfo {author} {\bibfnamefont
  {A.}~\bibnamefont {Werner}}, \bibinfo {author} {\bibfnamefont
  {C.}~\bibnamefont {N{\"u}hrenberg}}, \bibinfo {author} {\bibfnamefont
  {S.}~\bibnamefont {Ohdachi}}, \bibinfo {author} {\bibfnamefont
  {Y.}~\bibnamefont {Suzuki}}, \bibinfo {author} {\bibfnamefont
  {H.}~\bibnamefont {Yamada}}, \bibinfo {author} {\bibnamefont {the
  W7-AS~team}}, \ and\ \bibinfo {author} {\bibnamefont {the LHD~team}},\
  }\href@noop {} {\bibfield  {journal} {\bibinfo  {journal} {Fusion Science and
  Technology}\ }\textbf {\bibinfo {volume} {50}},\ \bibinfo {pages} {158}
  (\bibinfo {year} {2006})}\BibitemShut {NoStop}%
\bibitem [{\citenamefont {Moroz}(1996)}]{moroz1996}%
  \BibitemOpen
  \bibfield  {author} {\bibinfo {author} {\bibfnamefont {P.~E.}\ \bibnamefont
  {Moroz}},\ }\href@noop {} {\bibfield  {journal} {\bibinfo  {journal} {Physics
  of Plasmas}\ }\textbf {\bibinfo {volume} {3}},\ \bibinfo {pages} {3055}
  (\bibinfo {year} {1996})}\BibitemShut {NoStop}%
\bibitem [{\citenamefont {Shaing}\ \emph {et~al.}(1989)\citenamefont {Shaing},
  \citenamefont {Crume}, \citenamefont {Tolliver}, \citenamefont {Hirshman},\
  and\ \citenamefont {van Rij}}]{shaing1989}%
  \BibitemOpen
  \bibfield  {author} {\bibinfo {author} {\bibfnamefont {K.~C.}\ \bibnamefont
  {Shaing}}, \bibinfo {author} {\bibfnamefont {E.~C.}\ \bibnamefont {Crume}},
  \bibinfo {author} {\bibfnamefont {J.~S.}\ \bibnamefont {Tolliver}}, \bibinfo
  {author} {\bibfnamefont {S.~P.}\ \bibnamefont {Hirshman}}, \ and\ \bibinfo
  {author} {\bibfnamefont {W.~I.}\ \bibnamefont {van Rij}},\ }\href@noop {}
  {\bibfield  {journal} {\bibinfo  {journal} {Physics of Fluids B}\ }\textbf
  {\bibinfo {volume} {1}},\ \bibinfo {pages} {148} (\bibinfo {year}
  {1989})}\BibitemShut {NoStop}%
\bibitem [{\citenamefont {Post}\ \emph {et~al.}(1977)\citenamefont {Post},
  \citenamefont {Jensen}, \citenamefont {Tarter}, \citenamefont {Grasberger},\
  and\ \citenamefont {Lokke}}]{post1977}%
  \BibitemOpen
  \bibfield  {author} {\bibinfo {author} {\bibfnamefont {D.~E.}\ \bibnamefont
  {Post}}, \bibinfo {author} {\bibfnamefont {R.~V.}\ \bibnamefont {Jensen}},
  \bibinfo {author} {\bibfnamefont {C.~B.}\ \bibnamefont {Tarter}}, \bibinfo
  {author} {\bibfnamefont {W.~H.}\ \bibnamefont {Grasberger}}, \ and\ \bibinfo
  {author} {\bibfnamefont {W.~A.}\ \bibnamefont {Lokke}},\ }in\ \href@noop {}
  {\emph {\bibinfo {booktitle} {Atomic Data and Nuclear Data Tables}}},\
  Vol.~\bibinfo {volume} {20}\ (\bibinfo  {publisher} {Academic Press, Inc.},\
  \bibinfo {year} {1977})\ pp.\ \bibinfo {pages} {397--439}\BibitemShut
  {NoStop}%
\bibitem [{\citenamefont {Greene}\ and\ \citenamefont
  {Chance}(1981)}]{greene1981}%
  \BibitemOpen
  \bibfield  {author} {\bibinfo {author} {\bibfnamefont {J.~M.}\ \bibnamefont
  {Greene}}\ and\ \bibinfo {author} {\bibfnamefont {M.~S.}\ \bibnamefont
  {Chance}},\ }\href@noop {} {\bibfield  {journal} {\bibinfo  {journal}
  {Nuclear Fusion}\ }\textbf {\bibinfo {volume} {21}},\ \bibinfo {pages} {453}
  (\bibinfo {year} {1981})}\BibitemShut {NoStop}%
\bibitem [{\citenamefont {Gryaznevich}\ \emph {et~al.}(1998)\citenamefont
  {Gryaznevich}, \citenamefont {Akers}, \citenamefont {Carolan}, \citenamefont
  {Conway}, \citenamefont {Gates}, \citenamefont {Field}, \citenamefont
  {Hender}, \citenamefont {Jenkins}, \citenamefont {Martin}, \citenamefont
  {Nightingale}, \citenamefont {Ribeiro}, \citenamefont {Robinson},
  \citenamefont {Sykes}, \citenamefont {Tournianski}, \citenamefont {Valovi{\v
  c}},\ and\ \citenamefont {Walsh}}]{gryaznevich1998}%
  \BibitemOpen
  \bibfield  {author} {\bibinfo {author} {\bibfnamefont {M.}~\bibnamefont
  {Gryaznevich}}, \bibinfo {author} {\bibfnamefont {R.}~\bibnamefont {Akers}},
  \bibinfo {author} {\bibfnamefont {P.~G.}\ \bibnamefont {Carolan}}, \bibinfo
  {author} {\bibfnamefont {N.~J.}\ \bibnamefont {Conway}}, \bibinfo {author}
  {\bibfnamefont {D.}~\bibnamefont {Gates}}, \bibinfo {author} {\bibfnamefont
  {A.~R.}\ \bibnamefont {Field}}, \bibinfo {author} {\bibfnamefont {T.~C.}\
  \bibnamefont {Hender}}, \bibinfo {author} {\bibfnamefont {I.}~\bibnamefont
  {Jenkins}}, \bibinfo {author} {\bibfnamefont {R.}~\bibnamefont {Martin}},
  \bibinfo {author} {\bibfnamefont {M.~P.~S.}\ \bibnamefont {Nightingale}},
  \bibinfo {author} {\bibfnamefont {C.}~\bibnamefont {Ribeiro}}, \bibinfo
  {author} {\bibfnamefont {D.~C.}\ \bibnamefont {Robinson}}, \bibinfo {author}
  {\bibfnamefont {A.}~\bibnamefont {Sykes}}, \bibinfo {author} {\bibfnamefont
  {M.}~\bibnamefont {Tournianski}}, \bibinfo {author} {\bibfnamefont
  {M.}~\bibnamefont {Valovi{\v c}}}, \ and\ \bibinfo {author} {\bibfnamefont
  {M.~J.}\ \bibnamefont {Walsh}},\ }\href@noop {} {\bibfield  {journal}
  {\bibinfo  {journal} {Physical Review Letters}\ }\textbf {\bibinfo {volume}
  {80}},\ \bibinfo {pages} {3972} (\bibinfo {year} {1998})}\BibitemShut
  {NoStop}%
\bibitem [{\citenamefont {Chen}(1986)}]{chen1986}%
  \BibitemOpen
  \bibfield  {author} {\bibinfo {author} {\bibfnamefont {F.~F.}\ \bibnamefont
  {Chen}},\ }\href@noop {} {\emph {\bibinfo {title} {Introduction to Plasma
  Physics and Controlled Fusion}}},\ \bibinfo {edition} {2nd}\ ed.\ (\bibinfo
  {publisher} {Plenum Press},\ \bibinfo {address} {New York, NY},\ \bibinfo
  {year} {1986})\BibitemShut {NoStop}%
\bibitem [{\citenamefont {Yamada}\ \emph {et~al.}(2005)\citenamefont {Yamada},
  \citenamefont {Harris}, \citenamefont {Dinklage}, \citenamefont {Ascasibar},
  \citenamefont {Sano}, \citenamefont {Okamura}, \citenamefont {Talmadge},
  \citenamefont {Stroth}, \citenamefont {Kus}, \citenamefont {Murakami},
  \citenamefont {Yokoyama}, \citenamefont {Beidler}, \citenamefont {Tribaldos},
  \citenamefont {Watanabe},\ and\ \citenamefont {Suzuki}}]{yamada2005}%
  \BibitemOpen
  \bibfield  {author} {\bibinfo {author} {\bibfnamefont {H.}~\bibnamefont
  {Yamada}}, \bibinfo {author} {\bibfnamefont {J.~H.}\ \bibnamefont {Harris}},
  \bibinfo {author} {\bibfnamefont {A.}~\bibnamefont {Dinklage}}, \bibinfo
  {author} {\bibfnamefont {E.}~\bibnamefont {Ascasibar}}, \bibinfo {author}
  {\bibfnamefont {F.}~\bibnamefont {Sano}}, \bibinfo {author} {\bibfnamefont
  {S.}~\bibnamefont {Okamura}}, \bibinfo {author} {\bibfnamefont
  {J.}~\bibnamefont {Talmadge}}, \bibinfo {author} {\bibfnamefont
  {U.}~\bibnamefont {Stroth}}, \bibinfo {author} {\bibfnamefont
  {A.}~\bibnamefont {Kus}}, \bibinfo {author} {\bibfnamefont {S.}~\bibnamefont
  {Murakami}}, \bibinfo {author} {\bibfnamefont {M.}~\bibnamefont {Yokoyama}},
  \bibinfo {author} {\bibfnamefont {C.~D.}\ \bibnamefont {Beidler}}, \bibinfo
  {author} {\bibfnamefont {V.}~\bibnamefont {Tribaldos}}, \bibinfo {author}
  {\bibfnamefont {K.~Y.}\ \bibnamefont {Watanabe}}, \ and\ \bibinfo {author}
  {\bibfnamefont {Y.}~\bibnamefont {Suzuki}},\ }\href@noop {} {\bibfield
  {journal} {\bibinfo  {journal} {Nuclear Fusion}\ }\textbf {\bibinfo {volume}
  {45}},\ \bibinfo {pages} {1684} (\bibinfo {year} {2005})}\BibitemShut
  {NoStop}%
\bibitem [{\citenamefont {Seiwald}\ \emph {et~al.}(2007)\citenamefont
  {Seiwald}, \citenamefont {Nemov}, \citenamefont {Pedersen},\ and\
  \citenamefont {Kernbichler}}]{seiwald2007}%
  \BibitemOpen
  \bibfield  {author} {\bibinfo {author} {\bibfnamefont {B.}~\bibnamefont
  {Seiwald}}, \bibinfo {author} {\bibfnamefont {V.~V.}\ \bibnamefont {Nemov}},
  \bibinfo {author} {\bibfnamefont {T.~S.}\ \bibnamefont {Pedersen}}, \ and\
  \bibinfo {author} {\bibfnamefont {W.}~\bibnamefont {Kernbichler}},\
  }\href@noop {} {\bibfield  {journal} {\bibinfo  {journal} {Plasma Physics and
  Controlled Fusion}\ }\textbf {\bibinfo {volume} {49}},\ \bibinfo {pages}
  {2063} (\bibinfo {year} {2007})}\BibitemShut {NoStop}%
\bibitem [{\citenamefont {Sudo}\ \emph {et~al.}(1990)\citenamefont {Sudo},
  \citenamefont {Takeiri}, \citenamefont {Zushi}, \citenamefont {Sano},
  \citenamefont {Itoh}, \citenamefont {Kondo},\ and\ \citenamefont
  {IIyoshi}}]{sudo1990}%
  \BibitemOpen
  \bibfield  {author} {\bibinfo {author} {\bibfnamefont {S.}~\bibnamefont
  {Sudo}}, \bibinfo {author} {\bibfnamefont {Y.}~\bibnamefont {Takeiri}},
  \bibinfo {author} {\bibfnamefont {H.}~\bibnamefont {Zushi}}, \bibinfo
  {author} {\bibfnamefont {F.}~\bibnamefont {Sano}}, \bibinfo {author}
  {\bibfnamefont {K.}~\bibnamefont {Itoh}}, \bibinfo {author} {\bibfnamefont
  {K.}~\bibnamefont {Kondo}}, \ and\ \bibinfo {author} {\bibfnamefont
  {A.}~\bibnamefont {IIyoshi}},\ }\href@noop {} {\bibfield  {journal} {\bibinfo
   {journal} {Nuclear Fusion}\ }\textbf {\bibinfo {volume} {30}},\ \bibinfo
  {pages} {11} (\bibinfo {year} {1990})}\BibitemShut {NoStop}%
\bibitem [{\citenamefont {Giannone}\ \emph {et~al.}(2003)\citenamefont
  {Giannone}, \citenamefont {Brakel}, \citenamefont {Burhenn}, \citenamefont
  {Ehmler}, \citenamefont {Feng}, \citenamefont {Grigull}, \citenamefont
  {McCormick}, \citenamefont {Wagner}, \citenamefont {Baldzuhn}, \citenamefont
  {Igitkhanov}, \citenamefont {Knauer}, \citenamefont {Nishimura},
  \citenamefont {Pasch}, \citenamefont {Peterson}, \citenamefont
  {Ramasubramanian}, \citenamefont {Rust}, \citenamefont {Weller},
  \citenamefont {Werner},\ and\ \citenamefont {{the W7-AS
  Team}}}]{giannone2003}%
  \BibitemOpen
  \bibfield  {author} {\bibinfo {author} {\bibfnamefont {L.}~\bibnamefont
  {Giannone}}, \bibinfo {author} {\bibfnamefont {R.}~\bibnamefont {Brakel}},
  \bibinfo {author} {\bibfnamefont {R.}~\bibnamefont {Burhenn}}, \bibinfo
  {author} {\bibfnamefont {H.}~\bibnamefont {Ehmler}}, \bibinfo {author}
  {\bibfnamefont {Y.}~\bibnamefont {Feng}}, \bibinfo {author} {\bibfnamefont
  {P.}~\bibnamefont {Grigull}}, \bibinfo {author} {\bibfnamefont
  {K.}~\bibnamefont {McCormick}}, \bibinfo {author} {\bibfnamefont
  {F.}~\bibnamefont {Wagner}}, \bibinfo {author} {\bibfnamefont
  {J.}~\bibnamefont {Baldzuhn}}, \bibinfo {author} {\bibfnamefont
  {Y.}~\bibnamefont {Igitkhanov}}, \bibinfo {author} {\bibfnamefont
  {J.}~\bibnamefont {Knauer}}, \bibinfo {author} {\bibfnamefont
  {K.}~\bibnamefont {Nishimura}}, \bibinfo {author} {\bibfnamefont
  {E.}~\bibnamefont {Pasch}}, \bibinfo {author} {\bibfnamefont {B.~J.}\
  \bibnamefont {Peterson}}, \bibinfo {author} {\bibfnamefont {N.}~\bibnamefont
  {Ramasubramanian}}, \bibinfo {author} {\bibfnamefont {N.}~\bibnamefont
  {Rust}}, \bibinfo {author} {\bibfnamefont {A.}~\bibnamefont {Weller}},
  \bibinfo {author} {\bibfnamefont {A.}~\bibnamefont {Werner}}, \ and\ \bibinfo
  {author} {\bibnamefont {{the W7-AS Team}}},\ }\href@noop {} {\bibfield
  {journal} {\bibinfo  {journal} {Plasma Physics and Controlled Fusion}\
  }\textbf {\bibinfo {volume} {45}},\ \bibinfo {pages} {1713} (\bibinfo {year}
  {2003})}\BibitemShut {NoStop}%
\end{thebibliography}
\end{document}